\documentclass[journal=jacsat,manuscript=article]{achemso}
\SectionNumbersOn
\usepackage[version=3]{mhchem} 
\usepackage{amsmath,mathtools}
\usepackage{graphicx}
\graphicspath{ {./images/} }
\usepackage[font=small,labelfont=bf]{caption}
\usepackage[font=small,labelfont=bf]{subcaption}
\usepackage[dvipsnames]{xcolor}
\usepackage{booktabs,siunitx}
\usepackage{threeparttable}
\usepackage{hyperref}
\usepackage{wrapfig}

\newcommand{\bra}[1]{\ensuremath{\left\langle#1\right|}}
\newcommand{\ket}[1]{\ensuremath{\left|#1\right\rangle}}
\newcommand{\bracket}[2]{\ensuremath{\left\langle #1 \middle| #2 \right\rangle}}
\newcommand{\bigket}[1]{\ensuremath{\Big\vert #1 \Big\rangle}}
\newcommand{\ketbra}[2]{\ensuremath{\left|#1\right\rangle\left\langle#2\right|}}

\author{Junhan Chen}
\affiliation[Upenn]
{Department of Chemistry, University of Pennsylvania, Philadelphia, Pennsylvania 19104, USA}
\author{Zuxin Jin}
\affiliation[Upenn]
{Department of Chemistry, University of Pennsylvania, Philadelphia, Pennsylvania 19104, USA}
\author{Wenjie Dou}
\affiliation[UCB]
{Department of Chemistry, University of California Berkeley, Berkeley, California 94720, USA}
\author{Joseph Subotnik}
\email{subotnik@sas.upenn.edu}
\phone{+1 (215) 746-7078}
\affiliation[Upenn]
{Department of Chemistry, University of Pennsylvania, Philadelphia, Pennsylvania 19104, USA}

\title[An \textsf{achemso} demo]
  {Electronic Structure for Multielectronic Molecules Near a Metal Surface}

\abbreviations{IPOOs, IPVOs, OEOs, VEOs, OEs, VEs, NRG, MFT, UHF, HF}

\begin{document}
\begin{abstract}
  We analyze a model problem representing a multi-electronic molecule sitting on a metal surface.   
Working with a reduced configuration interaction Hamiltonian, we show that one can extract very accurate ground state wavefunctions
as compared with the numerical renormalization group theory (NRG) -- even in the limit of weak metal-molecule coupling strength
but strong intramolecular electron-electron repulsion.  Moreover, we extract what appear to be meaningful
excitation energies as well. Our findings should lay the groundwork for future {\em ab initio} studies of charge transfer processes and bond making/breaking processes on metal surfaces.
\end{abstract}

\newpage

\section{Introduction}

\hspace{2.5ex} Understanding the metal-molecule interface is essential for understanding heterogeneous catalysis. In order to understand macroscopically why heterogeneous catalysts enhance reaction rates and improve the production yield, we require a microscopic understanding of chemical reactions on an atomic scale. To that end, developing robust and atomistic quantum models of molecular processes that occur on metal surfaces, including electron-coupled adsorption \cite{bunermann2015electron} and electron-coupled vibration \cite{morin1992vibrational, huang2000vibrational}, is an important goal for modern theory. And in order to achieve such a goal, at least within the standard Born-Oppenheimer framework, the very first step is to solve the interfacial electronic structure problem.  If we can calculate potential energies that are accurate enough, then a host of dynamical approaches for the nuclear problem (Marcus theory\cite{marcus1956theory} and beyond\cite{hush1961adiabatic,levich1960adiabatic,cukier1998proton,jortner1998charge}) will be applicable and new dynamical techniques are still being developed \cite{dou2020nonadiabatic, shenvi2009nonadiabatic, dou2016broadened}.

Unfortunately, solving electronic structure problems is a very difficult task (in general), even for isolated molecules in the gas phase.  As is well known,  Hartree-Fock (HF) calculations show large discrepancies with experimental results even for isolated molecules \cite{dunning1989gaussian}: the H+H$_2$ reaction barrier \cite{baer1985theory,liu1984classical} , the dissociation energy for hydrogen fluoride \cite{nesbet1962approximate, herzberg1950molecular} , the ionization potential \cite{roothaan1963accurate, moore1949atomic} and the electron affinity \cite{sasaki1974configuration, elder1965electron} of the oxygen atom. Thus, even for small molecules, electron-electron correlation is important and of course expensive, scaling exponentially with the number of electrons. For large molecules, the situation is worse: one recent photochemistry study revealed that a correct treatment of electron correlation is needed to find the correct open-shell radical products (instead of closed-shell singlet
products) in the case of a bond-breaking reaction with carbenes and biradicals\cite{grafenstein2000can}.  Obviously, an accurate 
treatment of electronic correlation is essential for theory to match experiments even in the gas phase.

Now, if the state of affairs above (vis-à-vis electronic correlation for molecules in the gas phase) is unfortunate, the state of affairs in condensed matter physics is even worse. In the solid world, on-site electron-electron repulsion can result in metal-insulator transitions for narrow energy bands, e.g. the d-band in transition metals \cite{imada1998metal} and half-filling magic-angle graphene \cite{cao2018correlated}.  Within a solid, the electron correlation problem can couple together electronic states that are far apart not only in energy, but also in space, and perturbative treatments of electron correlation will often not be helpful. In the end, the computational cost needed to accurately solve the electronic structure problem in the condensed phase becomes simply immense and is motivating an enormous push today within the physics community.\cite{PhysRevResearch.2.032027,cevolani2018universal,PhysRevLett.121.067601,keshavarz2018electronic, knizia2012density}

With this background in mind, the theory of interfacial electronic structure (i.e. electronic structure for molecules on metal surfaces) lies somewhere in between the two extreme limits above.  
On the one hand, the interfacial problem has all of the difficulties described above as far as the electronic structure calculations of molecules.  To accurately describe a molecule on a metal surface, we require a sufficient treatment of static correlation (to describe bond breaking) as well as a sufficient treatment of dynamical correlation (to describe accurate, molecular orbital energies).  Moreover, when describing dynamic correlation, one must also take into account the orbital energies on the metal.  On the other hand, however, the interfacial electronic structure is easier than the condensed phase problem insofar as the fact that one can focus most of his/her attention on the molecule. For the most part, the static correlation problem is localized in space on the molecule (even if the dynamic correlation problem is spread out over the molecule and the metal).
To describe correlation in solids, one typically follows fermi liquid theory and uses DFT or some other effective mean-field theories. As a result, the interfacial problem is effectively an impurity problem, for which there is a significant literature
going back to the original Anderson model of a localized magnetic state in a sea of metallic electrons \cite{anderson1961localized}; 
The simplest one-site Anderson impurity model has been studied by a variety of impurity solvers including the numerical renormalization group (NRG) \cite{bulla2008numerical}, exact diagonalization (ED) \cite{fu2016numerical} and quantum monte carlo (QMC) \cite{gull2011continuous}.
More generally, solving the embedding problem in quantum chemistry has attracted a great deal of attention in recent years.\cite{wouters2016practical,knizia2012density,lee2019projection,bulik2014density,bulik2014electron,kluner2002periodic,sharifzadeh2008embedded,libisch2014embedded}

For our purposes, we are interested in coupled nuclear-electronic processes that occur at metal-molecule interfaces, especially
electron transfer processes and bond making or breaking processes. For such processes, with two stable configurations (e.g., donor and acceptor), we can certainly expect that static correlation effects will be essential, and one must go beyond mean-field theories \cite{newns1969self}. Within the quantum chemistry community, this line of thinking leads to different techniques in the literature.
\begin{enumerate}
    \item For processes that involve the charge character of the system, constrained DFT (CDFT) \cite{kaduk2012constrained,ma2020pycdft} is perhaps the simplest means to generate diabatic states and charge transfer excited states. This technique works extremely well in the limit of weak coupling (e.g. O$_2$ on Al(111)\cite{behler2007nonadiabatic} and benzene on Li(100)\cite{souza2013constrained}) but shows larger errors for strong coupling (e.g. N$_2$ on Ni(001) \cite{gavnholt2008delta}).  
    
    \item Beyond CDFT, there is of course a natural hierachy of 
    increasingly expensive wave-function techniques, including 
    multi-reference configuration interaction (MRCI) methods and/or multi-configurational self-consistent field (MCSCF). In particular, for problems with static correlation, the methods of choice today remain complete active space (CAS) \cite{roos1980complete, schmidt1998construction} methods. According to the definition of CAS, one usually chooses valence orbitals as the active space and the remaining inactive space refers to those orbitals which are either always occupied or always unoccupied. Here, a CAS(N,m,S) represents N active electrons in m active orbitals with total spin quantum number S (strictly speaking, S is equal to one-half of the number of singly occupied orbitals). The number of configurations contained in a MCSCF wavefunction is given by the Weyl-Robinson formula \cite{schmidt1998construction, pauncz1995symmetric}:
    \begin{equation}
        \frac{2S+1}{m+1}\binom{m+1}{m-\frac{N}{2}-S}\binom{m+1}{\frac{N}{2}-S}
    \end{equation}
    When N and m are small, CAS methods are accurate and fast.  However, for larger molecules, with finite speed and memory capabilities, one cannot afford to include very many orbitals in the active space of a CASSCF calculation--even with advanced bookkeeping techniques\cite{szabo2012modern}. Moreover, due to the exponential scaling of the number of Slater determinants with the number of orbitals and electrons, the practical upper limit for CAS is about 24 electrons in 24 active orbitals. \cite{olsen2011casscf} (
   In the context of the density matrix renormalization group (DMRG) algorithm, the number of active orbitals can go as large as 100 \cite{chan2011density}.) As a result, for accurate energies, within the molecular community, one tries (if possible) to include dynamical correlation either by calculating a second-order perturbation correction (MR-PT, e.g. CASPT2 \cite{andersson1992second}), solving a configuration interaction (CI) problem with MCSCF wavefunctions as the reference states (e.g. CAS(2,2)CISD \cite{zgid2012truncated}), or even solving for a multi-reference couple-cluster solution (MR-CC, e.g. ACPF \cite{gdanitz1988averaged}).
    
    \item Lastly, historically, there have also been attempts to merge DFT with CI to recover the multiconfigurational character for {\em molecules} \cite{grimme1999combination}.
\end{enumerate}

Of the methods listed above, only CDFT has been applied to realistic metal surface, and with varying degrees of success \cite{kaduk2012constrained,ma2020pycdft,souza2013constrained,behler2007nonadiabatic}. For the truly multiconfigurational approaches, the computational cost is enormous and these techniques are not readily used to study interfacial electronic structure (though see recent work of Levine {\em et al} for an interesting CAS calculation of dangling bonds on a silicon cluster).\cite{peng2018dynamics}

At this point, it should be clear to the reader that strong approximations will be necessary in order to practically and robustly solve for the electronic structure of a molecule reacting on a metal surface, while capturing enough correlation energy for even qualitative (and ideally quantitative) accuracy. 
In order to achieve such a goal, in this paper, we will follow a three-pronged approach. First, we will run a standard SCF/DFT calculation
to allow for electrons to be delocalized; second, we will use a projection technique,  which is similar to the framework of Chan's density matrix embedding theory (DMET) \cite{wouters2016practical}, to generate molecular orbitals corresponding to a molecule on a metal surface; third, we will generate and diagonalize a configuration interaction Hamiltonian 
that will allow for multi-reference behavior while also yielding information about excited states.  For the present paper, we will restrict ourselves to a two-site impurity with electron repulsion (representing a molecule) coupled to a set of non-interacting fermions (representing a metal bath)--but in the future, we intend to apply the present approach
to {\em ab initio} (rather than model) calculations. For now, though,
in order to make sure that we recover accurate results for a model problem, we will compare all of our configuration interaction data against (exact) the numerical renormalization group theory (NRG)
\cite{bulla2008numerical}, which is expensive but possible for a small model Hamiltonian. In the end, our hope is that the present methodology (or some variants) should allow us to model accurately the dynamics of molecular charge transfer, bond-making or bond-breaking processes on a  metal surface.



This paper is organized as follows. In Sec. \ref{theory}, we introduce the two-site Anderson impurity Hamiltonian, which will serve as our model Hamiltonian (representing a many-electron molecule sitting on a metal surface). We further introduce the necessary projection
operators that are needed for constructing (effectively) an embedded set of orbitals on the impurity that interact with a metal surface.    In Sec. \ref{theory}(\ref{discussion1}), using the molecular orbitals just defined, we introduce a host of configuration interaction methods for approximating the total Hamiltonian (molecule plus metal).  In Sec. \ref{results},  we present results, demonstrating that
for many parameter regimes, one can invoke (with accuracy) a CI technique we label CI(N-1, N-1), which includes N-1 singly excited configurations plus N-1 doubly excited configurations in total. In Sec. \ref{discussion2}, we further analyze our data and, in particular, we investigate
 the regime whereby one appears to break a molecular bond on the metal surface. For this parameter regime, 
we find that achieving accuracy requires
 at least two more doubly excited configurations (leading to a CI(N-1, N+1) ansatz), whose meaning we discuss in detail.  
In Sec. \ref{conclusion}, we summarize our findings and give an outlook for prospective future applications
with realistic {\em ab initio} systems.

A word about notation is now appropriate and essential. Henceforward, we will refer
to ``impurities'' when discussing a molecule (sitting on a metal surface) and we will refer to a ``bath'' when discussing the metal.  The term ``universe'' will denote the impurity plus the metal.
When performing electronic structure calculations,
many different sets of orbitals can and will be constructed. In what follows below, we will represent 
the underlying atomic orbital basis for our calculation as $\left\{\chi_{\nu}\right\}$, where $\nu$ runs over all sites in the universe.  We will represent the canonical Kohn-Sham or HF orbitals (which are delocalized over both the impurity and the metal bath) as $\left\{\tilde{\psi}_i\right\}$. Greek indices $(\mu,\nu)$ strictly index atomic sites, whereas roman indices index delocalized orbitals. 
As usual, $i,j,k$ index occupied delocalized orbitals, whereas $a,b,c$ index virtual delocalized orbitals.
The calculation below
will rely on the construction of impurity-projected occupied orbitals (IPOOs) 
and impurity-projected virtual orbitals (IPVOs), which are referenced (respectively) 
as $\left|\phi^{occ}_{\nu}\right>$ and and $\left|\phi^{virt}_{\nu}\right>$.
Finally, the most important set of orbitals constructed below will be those orbitals that span
the occupied canonical space, but have been separated into impurity and bath components; these orbitals
will be labeled $\left\{\left|\psi_1\right>, \left|\psi_2\right>, \ldots,\left|\psi_{h-1}\right> , \left|\psi_h\right>  \right\}$. Similarly orbitals will also be constructed for the virtual space:
$\left\{\left|\psi_l\right>, \left|\psi_{l+1}\right>,  \ldots, \left|\psi_N\right>  \right\}$.
\cite{Inprincipleinfinite}
Lastly, throughout this manuscript, we will attempt to avoid using the common phrase ``molecular orbitals'', 
which could easily refer to several of the orbital sets listed above.

\section{Theory}
\label{theory}
\subsection{Two-Site Anderson Impurity Model}
\indent For the present manuscript, our model Hamiltonian of choice will be the two-site Anderson impurity model. 
Within a second quantized representation, the Hamiltonian for the universe can be written as:
 \begin{equation}
 \begin{aligned}
     \hat{H}&=\epsilon_d\sum_\sigma d_{1\sigma}^{\dagger}d_{1\sigma}+(\epsilon_d+\Delta \epsilon_d)\sum_\sigma d_{2\sigma}^{\dagger}d_{2\sigma}+t_d\sum_\sigma(d_{1\sigma}^{\dagger}d_{2\sigma}+d_{2\sigma}^{\dagger}d_{1\sigma})\\
     +&U(d_{1\uparrow}^{\dagger}d_{1\uparrow}d_{1\downarrow}^{\dagger}d_{1\downarrow}+d_{2\uparrow}^{\dagger}d_{2\uparrow}d_{2\downarrow}^{\dagger}d_{2\downarrow})
     + \sum_{k\sigma}\epsilon_{k\sigma}c_{k\sigma}^{\dagger}c_{k\sigma}+\sum_{k\sigma}V(d_{1\sigma}^{\dagger}c_{k\sigma}+c_{k\sigma}^{\dagger}d_{1\sigma})
     \end{aligned}
 \end{equation}

The universe's Hamiltonian can be separated into two parts: the one-electron core Hamiltonian and the two-electron term:
\begin{equation}
    \hat{H} \equiv \hat{H}_{core} + \hat{\Pi}
\end{equation}
where,
\begin{equation}
\begin{aligned}
    \hat{H}_{core}&=\epsilon_d\sum_\sigma d_{1\sigma}^{\dagger}d_{1\sigma}+(\epsilon_d+\Delta \epsilon_d)\sum_\sigma d_{2\sigma}^{\dagger}d_{2\sigma}+t_d\sum_\sigma(d_{1\sigma}^{\dagger}d_{2\sigma}+d_{2\sigma}^{\dagger}d_{1\sigma})\\
     &+ \sum_{k\sigma}\epsilon_{k\sigma}c_{k\sigma}^{\dagger}c_{k\sigma}+\sum_{k\sigma}V(d_{1\sigma}^{\dagger}c_{k\sigma}+c_{k\sigma}^{\dagger}d_{1\sigma})
\end{aligned}
\label{eq:Hcore}
\end{equation}

\begin{equation}
    \hat{\Pi}=U(d_{1\uparrow}^{\dagger}d_{1\uparrow}d_{1\downarrow}^{\dagger}d_{1\downarrow}+d_{2\uparrow}^{\dagger}d_{2\uparrow}d_{2\downarrow}^{\dagger}d_{2\downarrow})
\end{equation}

\newpage

\begin{wrapfigure}{r}{0.5\textwidth}
  \centering
  \includegraphics[width=0.48\textwidth]{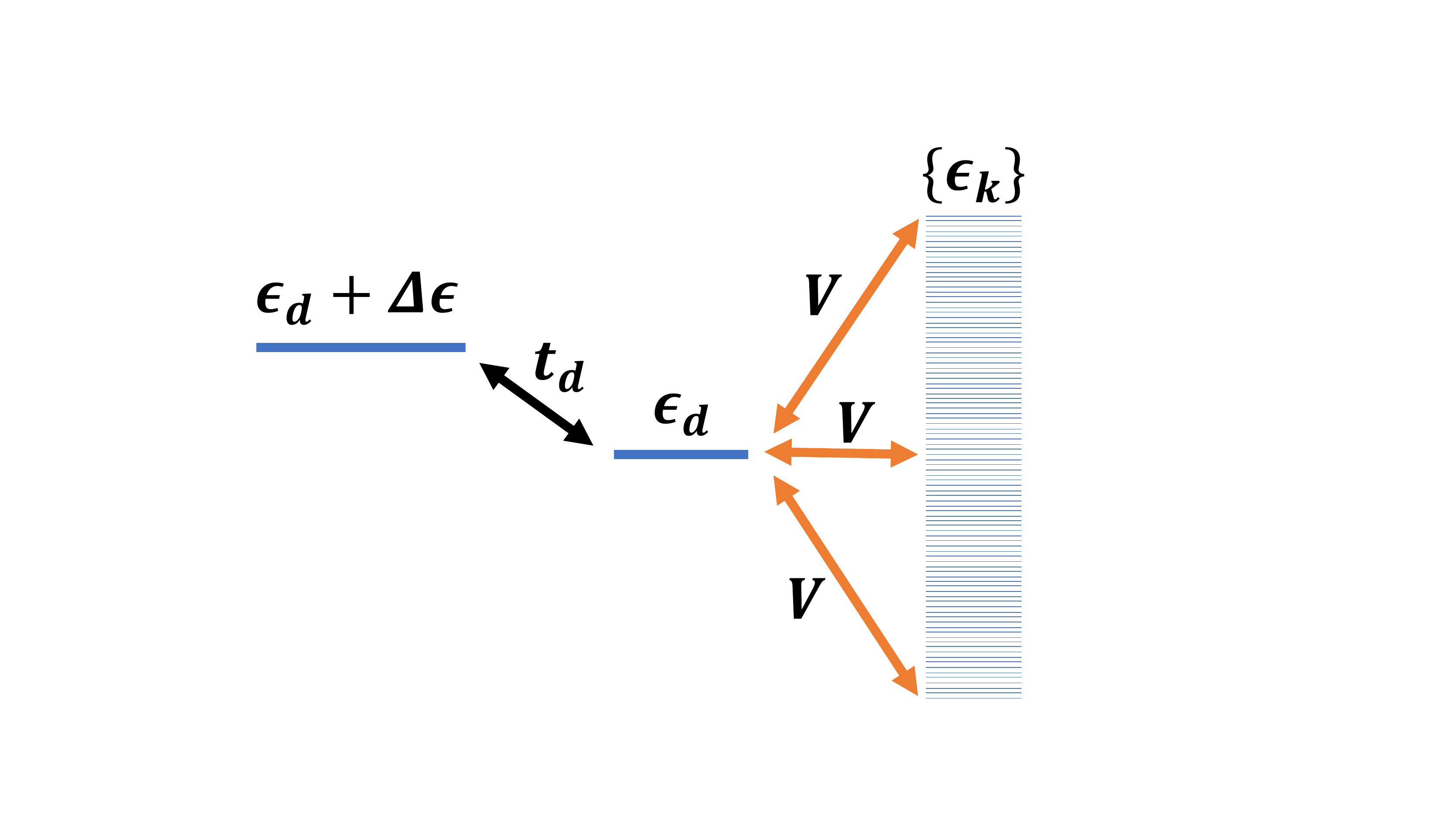}
  \caption{A schematic figure for the two-site Anderson impurity Hamiltonian.}
  \label{fig:hamiltonian}
\end{wrapfigure}
This Hamiltonian can be visualized as in Fig. \ref{fig:hamiltonian}.
The creation and annihilation operators $\{\hat{d}^\dagger,\hat{d}\}$ refer to impurity atomic orbitals (which should represent
a molecule on a surface), the operators $\{\hat{c}^\dagger,\hat{c}\}$ refer to bath (metal surface) atomic orbitals, and $\sigma$ 
refers to an electron spin.
$\epsilon_d$ and $\epsilon_d+\Delta \epsilon_d$ are ionization energies for the impurity site 1 and the site 2. $t_d$ is the hopping parameter between the site 1 and the site 2, $U$ represents the on-site coulomb repulsion for the impurity. $\epsilon_{k}$ represents the energy of the free-electron metallic orbital with momentum $k$, while $V$ represents the hybridization between the impurity site 1 and the metal bath. \cite{parameters}

For the present manuscript, in order to compare our results with (exact) the numerical renormalization group theory (NRG) results,
we will make the wide band approximation, whereby we assume that the hybridization 
width $\Gamma(\epsilon)=2\pi\sum_k |V|^2\delta(\epsilon-\epsilon_k)$ is independent of energy, i.e. $\Gamma(\epsilon)= \Gamma$. For all calculations below, we set $\Gamma = 0.01$ and $U=10\Gamma=0.1$. The band width for the bath is set to 0.8 hartrees, ranging from -0.4 to 0.4. 800 bath states are evenly distributed inside this energy window so that the energy spacing for the bath states is $\Delta E=0.001$ (i.e. the density of the bath states is $\rho(\epsilon) = 1000$). Extending the present method to go beyond the wide-band approximation is straightforward (and has already been implemented on our end).

\subsection{Choosing an  Orbital Active Space for the Molecular Impurity} 
In order to build up a configuration interaction Hamiltonian for a molecular impurity on a metal surface, we will require a set of 
impurity orbitals that we can associate as  belonging to the impurity. Although many projection schemes have been developed over the past
 several years \cite{marzari1997maximally, wouters2016practical}, we will choose an approach similar to 
 the projected wannier functions algorithm. \cite{marzari2012maximally}

We begin by performing a standard mean-field theory (MFT) calculation. The resulting MFT eigenstates $\{\tilde{\psi}_p\}$ are delocalized wavefunctions and come in two flavors: occupied and virtual. 
The projection operators for each subspace are $\hat{P}_{occ}$ and $\hat{P}_{vir}$:
\begin{equation}
    \hat{P}_{occ}=\sum_{i\in{occ}}\ket{\tilde{\psi}_i}\bra{\tilde{\psi}_i}
\end{equation}
\begin{equation}
        \hat{P}_{virt}=\hat{I}-\hat{P}_{occ}=\sum_{a\in{virt}}\ket{\tilde{\psi}_a}\bra{\tilde{\psi}_a}
\end{equation}

Next, we project the impurity atomic orbitals $\{\ket{\chi_{\nu}}^{imp}\}_{\nu=1,2}$ into the occupied and virtual subspaces
as follows:
\begin{equation}
    \ket{\chi_{\nu}}^{imp}=\underbrace{\hat{P}_{occ}\ket{\chi_{\nu}}^{imp}}_{\ket{\tilde{\phi}_{\nu}^{occ}}}+\underbrace{\hat{P}_{virt}\ket{\chi_{\nu}}^{imp}}_{\ket{\tilde\phi_{\nu}^{virt}}}
\end{equation}
Here, the projected functions are:
\begin{equation*}
    \begin{aligned}
    \ket{\tilde{\phi}_{\nu}^{occ}}=\sum_{i\in{occ}}\ket{\tilde{\psi}_{i}}\bracket{\tilde{\psi}_{i}}{\chi_{\nu}}^{imp}
    \\
    \ket{\tilde{\phi}_{\nu}^{virt}}=\sum_{a\in{virt}}\ket{\tilde{\psi}_{a}}\bracket{\tilde{\psi}_{a}}{\chi_{\nu}}^{imp}
    \end{aligned}
\end{equation*}
Note that, if we had a more complicated Hamiltonian representing a larger molecular system, we would have more than two atomic sites -- and yet the present formalism can be extended in an obvious fashion.

Lastly, the projected orbitals $\{\left|\tilde{\phi}_{\nu}^{occ}\right>\}_{\nu=1,2}$ and $\{\tilde{\phi}_{\nu}^{virt}\}_{\nu=1,2}$ are orthogonal in the sense that all occupied and virtual basis functions satisfy:$\langle\tilde{\phi}_{\mu}^{occ}\mid\tilde{\phi}_{\nu}^{virt}\rangle=0$. However, these functions are not orthogonal in the sense that:$\langle\tilde{\phi}_{\mu}^{occ}\mid\tilde{\phi}_{\nu}^{occ}\rangle\neq \delta_{\mu \nu}$ and $\langle\tilde{\phi}_{\mu}^{virt}\mid\tilde{\phi}_{\nu}^{virt}\rangle\neq \delta_{\mu \nu}$. Nevertheless, one can easily recover an orthonormal basis
by performing a Löwdin orthogonalization \cite{szabo2012modern}, yielding $\{\left|\phi_{\nu}^{occ}\right>\}_{\nu=1,2}$ and $\{\phi_{\nu}^{virt}\}_{\nu=1,2}$.

The steps above can be summarized mathematically (in the precise language of Ref. \cite{marzari2012maximally}) as follows:

\begin{enumerate}
\item Compute a matrix of inner products $(A_{occ})_{i\nu}=\bracket{\tilde{\psi}_{i}}{\chi_{\nu}}^{imp}$. Then the projection can be written as:
\begin{equation}
    \ket{\tilde{\phi}_{\nu}^{occ}}=\sum_{i\in{occ}}\ket{\tilde{\psi}_{i}}(A_{occ})_{i\nu}
\end{equation}

\item Compute the overlap matrix $(S_{occ})_{\mu\nu}=(A_{occ}^{\dagger}A_{occ})_{\mu\nu}$ 

\item Construct the Löwdin-orthogonalized impurity-projected occupied orbitals (IPOOs):
\begin{equation}
    \ket{\phi_{\nu}^{occ}}=\sum_{\mu}\ket{\tilde{\phi}_{\mu}^{occ}}(S_{occ}^{-1/2})_{\mu\nu}
    = \sum_{i\in{occ}}\ket{\tilde{\psi}_{i}}(A_{occ}S_{occ}^{-1/2})_{i\nu}
    \label{eq:ipoos}
\end{equation}

\end{enumerate}

Note that the quantity  $A_{occ}S_{occ}^{-1/2}$ in Eq. \ref{eq:ipoos}
is a unitary transformation.
After all, according to a singular value decomposition,
\begin{equation*}
\begin{aligned}
    (A_{occ})_{k\nu}&=\sum_{p=1}^2U_{kp}\lambda_p V^{\dagger}_{p\nu} \\
    (S_{occ})_{\mu\nu}=(A_{occ}^{\dagger}A_{occ})_{\mu\nu}&=\sum_{p=1}^2V_{\mu p}\lambda_p^2 V^{\dagger}_{p\nu} \\
    \rightarrow A_{occ}S_{occ}^{-1/2}&=UV^{\dagger}
    \end{aligned}
\end{equation*}


\subsection{Constructing Frontier Orbitals by Minimization of the Energy of a Double Excitation}
The IPOOs $\{\phi_{\nu}^{occ}\}_{\nu=1,2}$ and IPVOs $\{\phi_{\nu}^{virt}\}_{\nu=1,2}$ 
form an active subspace of orbitals for the impurity within the context of the two-site Hamiltonian considered here. More generally, one would like to work with
impurities (or really molecules) with many, many electrons. And so, in order to make progress with any form of electron-electron
correlation, we will need to construct HOMO and LUMO orbitals for the impurity.
To that end, we will roughly follow the approach in Ref. \cite{teh2019simplest}. This approach can be made very clear (and explicit) using the current simple model, with only two sites.

We begin by rotating the projected orbitals:
\begin{equation*}
\begin{aligned}
\begin{pmatrix}
    \ket{\psi_{h-1}} & \ket{\psi_h}
\end{pmatrix}
&=
\begin{pmatrix}
    \ket{\phi_{1}^{occ}} & \ket{\phi_{2}^{occ}} 
\end{pmatrix}
\begin{pmatrix}
    -sin(\theta_1) & cos(\theta_1) \\
    cos(\theta_1) & sin(\theta_1)
\end{pmatrix}
\\
\begin{pmatrix}
    \ket{\psi_{l}} & \ket{\psi_{l+1}}
\end{pmatrix}
&=
\begin{pmatrix}
    \ket{\phi_{1}^{virt}} & \ket{\phi_{2}^{virt}} 
\end{pmatrix}
\begin{pmatrix}
    cos(\theta_2) & -sin(\theta_2) \\
    sin(\theta_2) & cos(\theta_2)
\end{pmatrix}
\end{aligned}
\end{equation*} 
The premise of Ref. \cite{teh2019simplest} is to pick the angles $\theta_1$ and $\theta_2$ above
(and hence optimized orbitals $\{\ket{\psi_{h-1}},\ket{\psi_h}\}$ and $\{\ket{\psi_{l}},\ket{\psi_{l+1}}\}$)
by minimizing the energy for the doubly excited configuration: $\ket{\Psi_{h\bar{h}}^{l\bar{l}}}$. Explicitly, the energy
for this doubly excited configuration is (assuming a closed-shell restricted set of orbitals):
\begin{equation}
\begin{aligned}
    E_{double}&=2\mbox{tr}\left(\hat{H}_{core} \hat{P}_{occ}\right)-2(\psi_h|\hat{H}_{core}|\psi_h)+2(\psi_l|\hat{H}_{core}|\psi_l)\\
    &+U\left[(\langle \hat{n}_{1} \rangle-\bra{\psi_h}\hat{n}_1\ket{\psi_h}+\bra{\psi_l}\hat{n}_1\ket{\psi_l})^2
    +(\langle \hat{n}_{2} \rangle-\bra{\psi_h}\hat{n}_2\ket{\psi_h}+\bra{\psi_l}\hat{n}_2\ket{\psi_l})^2\right]\\
    &=E_{HF}-2F_{hh}+2F_{ll} \\
    &+U(\bra{\psi_h}\hat{n}_1\ket{\psi_h}-\bra{\psi_l}\hat{n}_1\ket{\psi_l})^2+U(\bra{\psi_h}\hat{n}_2\ket{\psi_h}-\bra{\psi_l}\hat{n}_2\ket{\psi_l})^2
    \end{aligned}
\end{equation}
Here, $E_{HF}$ is the Hartree-Fock ground state energy
\begin{equation}
    E_{HF}=2\mbox{tr}\left(\hat{H}_{core}\hat{P}_{occ}\right)+U\left(\langle \hat{n}_{1} \rangle^2+\langle \hat{n}_{2} \rangle^2\right),
\end{equation}
$\hat{H}_{core}$ is the one-electron 
term in Eq. \ref{eq:Hcore} and $\hat{F}$ is the fock operator as constructed by a standard HF calculation:
\begin{equation}
    \hat{F}=\hat{H}_{core}+U\left(\hat{n}_{1} \langle \hat{n}_{1} \rangle+\hat{n}_{2} \langle \hat{n}_{2} \rangle\right)
\end{equation}
Lastly, $F_{hh} \equiv \left<\psi_h \middle| \hat{F} \middle| \psi_h\right>$ and
$F_{ll} \equiv \left<\psi_l \middle| \hat{F} \middle| \psi_l\right>$.

\subsection{Putting It All Together: A Complete Basis That Extrapolates The Optimized Orbitals}
Having constructed $\left\{\ket{\psi_{h-1}}, \ket{\psi_{h}}\right\}$, 
we can extend this two dimensional set of vectors to include $N_{occ}-2$ more functions 
so as to form a complete basis for the occupied space.  To do this in the most numerically 
stable fashion, we construct $\{\psi_i\}_{i=1,...,N_{occ}-2}$ (which label the bath) according to  
a standard canonical orthogonalization procedure\cite{szabo2012modern}. Namely, we first calculate the projector onto the reduced occupied space:
\begin{equation}
    S=\hat{P}_{occ}-\ketbra{\psi_{h-1}}{\psi_{h-1}}-\ketbra{\psi_h}{\psi_h}
\end{equation}

Next, we express $S$ in the basis of atomic orbitals $(N \times N)$
and then diagonalize it:
\begin{equation}
    s=V'SV
\end{equation}

If everything above is completely stable numerically, we should find that $S$ has 
$N_{vir} +2$ zero eigenvalues.  And even if numerical instabilities arise,
we can always just sort the resulting eigenvalues in descending
 order. In the end, we can  generate a coefficient matrix $\tilde{X}$:
\begin{equation}
    \tilde{X}=
    \begin{bmatrix}
    V_{1,1}/s_1^{1/2} & V_{1,2}/s_2^{1/2} & \dots & V_{1, N_o-2}/s_{N_o-2}^{1/2} \\
    V_{2,1}/s_1^{1/2} & V_{2,2}/s_2^{1/2} & \dots & V_{2, N_o-2}/s_{N_o-2}^{1/2} \\
    \vdots & \vdots & \ddots & \vdots \\
    V_{N,1}/s_1^{1/2} & V_{N,2}/s_2^{1/2} & \dots & V_{N, N_o-2}/s_{N_o-2}^{1/2}
    \end{bmatrix}
    \label{eq:X}
\end{equation}
that gives us a prescription for the complete occupied space:
\begin{eqnarray}
\ket{\psi_i} = \sum_\nu \tilde{X}_{\nu i} \chi_{\nu} ,\quad i=1,2,...,N_o-2
\end{eqnarray}
Note that the $\sqrt{s_i}$ factors in the denominators on the right hand side of Eq. \ref{eq:X} are included only to normalize the $\{\ket{\psi_i}\}$ functions.

In the end, the take-away message from this entire section is that we have constructed a 
complete set of occupied orbitals ($\ket{\psi_1}, \ket{\psi_2},...,\ket{\psi_{h-1}},\ket{\psi_h}$) whereby $\ket{\psi_{h-1}}$ and $\ket{\psi_h}$ can be associated with the impurity, and all other orbitals are associated with the bath. Of course, the same procedure can be done for the virtual space (where now we work with $\ket{\psi_l}$ and $\ket{\psi_{l+1}}$ instead of $\ket{\psi_{h-1}}$ and $\ket{\psi_{h}}$).
Henceforward, in the spirit of DMET \cite{wouters2016practical}, we will call these two sets of orbitals $\left\{ \psi_{h-1}, \psi_{h} \right\}$ and $\left\{ \psi_{l}, \psi_{l+1} \right\}$ occupied entangled orbitals(OEOs) and virtual entangled orbitals(VEOs), respectively. We will refer to the remaining two sets of orbitals $\left\{ \psi_{1}, \psi_{2}, ..., \psi_{h-2} \right\}$ and $\left\{\psi_{l+2}, ... \psi_N \right\}$ as occupied bath orbitals (OBOs)) and virtual bath orbitals (VBOs), respectively.


\subsection{Selecting a Configuration Interaction Basis}
\label{discussion1}

The goal of this paper is to establish and compare a set of different configuration interaction
methods for capturing the electronic structure of an impurity on a metal surface. To that end, 
in Table \ref{tbl:selective CI}, we list six possible CI Hamiltonians that will appear natural
to the seasoned quantum chemist/physicist. Our notation is as follows:

\begin{itemize}
\item CAS(2,2) [Complete Active Space(2,2)] represents all configurations with 2 electrons on 2 orbitals $\{\psi_h,\psi_l\}$.  

\item CI(X,Y) represents a selective configuration interaction Hamiltonian with only single
and double excitations: X is the number of singly excited configuration and Y is the number of doubly excited configuration.
\end{itemize}

Now, obviously, the notation CI(X,Y) is not unique: which X single and which Y double excitations should we include? Thus, in Table \ref{tbl:selective CI}, in the middle column, we list explicitly the configurations.  For example, CI(N$_{\textrm{ov}}$,1) includes N$_{\textrm{o}} \times$ N$_{\textrm{v}}$ singly excited configurations and one doubly excited configuration (which we called ``CIS-1D'' in Ref. \cite{teh2019simplest}). To better understand this table, note the following nomenclature conventions we have used:

\begin{enumerate}
\item $\{\bigket{\Phi_{\textrm{HF}}}\}$ denotes the Hartree-Fock ground state.

\item The subscript $i$ indexes all occupied orbitals, including occupied entangled orbitals $\{\psi_{h-1},\psi_h\}$ and occupied bath orbitals.

\item The subscript $a$ includes the virtual entangled orbital $\{\psi_{l+1}\}$ and all virtual bath orbitals, but excludes the virtual entangled orbital $\{\psi_l\}$ (in order to avoid double counting ($\ket{S_h^{a=l}}=\ket{S_{i=h}^l}$). 

\item The subscript $b$ in CI(N$_{\textrm{ov}}$,1) denotes all unoccupied orbitals, associate or not associated with the impurity. 

\item Every configuration is a singlet spin-adapted configuration: $\ket{S_h^l}=\frac{1}{\sqrt{2}}\left(\ket{\Phi_h^l}+\ket{\Phi_{\overline{h}}^{\overline{l}}}\right)$ and ${}^1\!\bigket{\Phi_{ih}^{ll}}=\frac{1}{\sqrt{2}}\left(\ket{\Phi_{i\overline{h}}^{l\overline{l}}}+\ket{\Phi_{\overline{i}h}^{\overline{l}l}}\right)$. For the case $i=h$, we set ${}^1\!\bigket{\Phi_{ih}^{ll}}=\bigket{\Phi_{h\overline{h}}^{l\overline{l}}}$.
\end{enumerate}

\begin{wrapfigure}{r}{0.5\textwidth}
  \centering
  \includegraphics[width=0.48\textwidth]{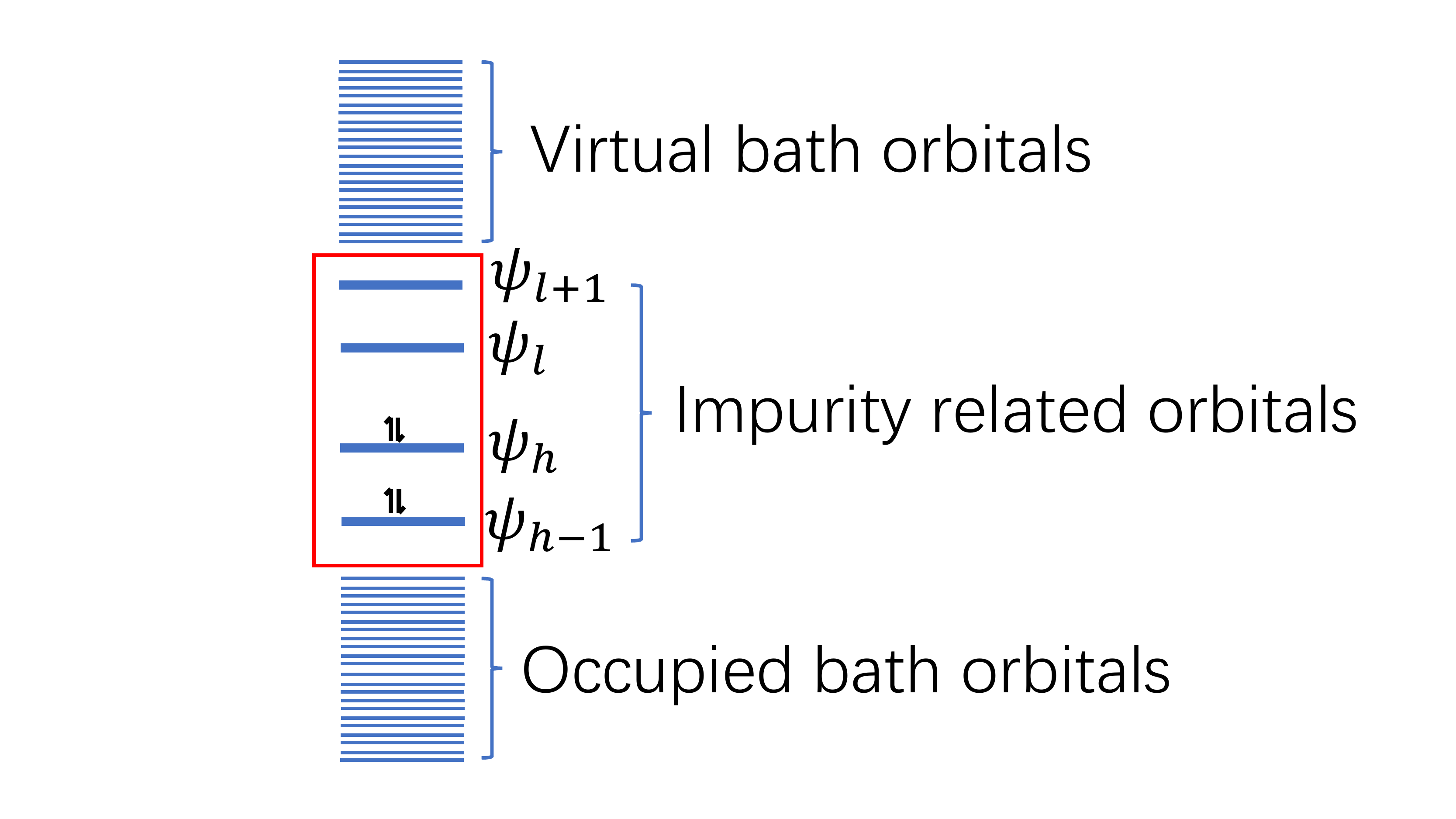}
  \caption{A schematic figure of the relevant orbitals\cite{orbitals}}
  \label{fig:active orbitals}
\end{wrapfigure}

On the right hand side of Table \ref{tbl:selective CI}, we list the total number of configurations for each approach. Here, N denotes the total number of orbitals, including N$_\textrm{o}$ number of occupied orbitals and N$_\textrm{v}$ number of virtual orbitals. As should be clear from the description above, we have 
constructed 2 occupied orbitals associated with the impurity, N$_\textrm{o}$-2 occupied orbitals associated with the bath, 2 virtual orbitals associated with the impurity, and N$_\textrm{v}$-2 virtual orbitals associated with the bath; see Fig. \ref{fig:active orbitals}.

Overall, the basic ansatz of the present paper is that, if we include enough configurations of relevance to the impurity, we should be able to recover reasonably accurate electronic structure.

\begin{table*} 
\centering
\caption{Selective CI Calculations}
\begin{threeparttable}
\begin{tabular}{ccc}
\toprule
Methods\tnote{*} & Configurations\tnote{**} & Number of Configurations (N=803) \\
  \midrule
\textcolor{BurntOrange} {CAS(2,2)}       & \textcolor{BurntOrange} { $\bigket{\Phi_{\textrm{HF}}}, \bigket{S_h^l}, \bigket{\Phi_{h\overline{h}}^{l\overline{l}}}$ } & \textcolor{BurntOrange} {3} \\
\textcolor{Dandelion} {CI(N-1,1)}  & \textcolor{Dandelion} {$\bigket{\Phi_{\textrm{HF}}}, \bigket{S_i^l}, \bigket{S_h^a}, \bigket{\Phi_{h\overline{h}}^{l\overline{l}}}$ }& \textcolor{Dandelion} {N+1 (804)}\\
\textcolor{Fuchsia} {CI(1,N-1)}       & \textcolor{Fuchsia} {$\bigket{\Phi_{\textrm{HF}}}, \bigket{S_h^l}, {}^1\!\bigket{\Phi_{ih}^{ll}}, {}^1\!\bigket{\Phi_{hh}^{al}}$ }& \textcolor{Fuchsia} {N+1 (804)}\\
\textcolor{red} {CI(N-1,N-1)} & \textcolor{red} {$\bigket{\Phi_{\textrm{HF}}}, \bigket{S_i^l}, \bigket{S_h^a}, {}^1\!\bigket{\Phi_{ih}^{ll}}, {}^1\!\bigket{\Phi_{hh}^{al}}$ }& \textcolor{red} {2N-1 (1605)}\\
\textcolor{blue} {CI(N-1,N+1)} & 
\begin{tabular}{@{}c@{}}
\textcolor{blue} {$\bigket{\Phi_{\textrm{HF}}}, \bigket{S_i^l}, \bigket{S_h^a}, {}^1\!\bigket{\Phi_{ih}^{ll}} {}^1\!\bigket{\Phi_{hh}^{al}}$} \\
\textcolor{blue} {$\bigket{\Phi_{h-1\overline{h-1}}^{l\overline{l}}}, \bigket{\Phi_{h\overline{h}}^{l+1\overline{l+1}}}$ }
\end{tabular}
& \textcolor{blue} {2N+1 (1607)}\\
\textcolor{OliveGreen} {CI(N$_{\textrm{ov}}$,1)}       & \textcolor{OliveGreen} {$\bigket{\Phi_{\textrm{HF}}}, \bigket{S_i^b}, \bigket{\Phi_{h\overline{h}}^{l\overline{l}}}$ }& \textcolor{OliveGreen} {N$_\textrm{o}$N$_\textrm{v}$+2 (161202)}\\
  \bottomrule
  \end{tabular}
      \begin{tablenotes}
      
      \footnotesize
      \item[*] CI(X,Y) represents X number of singly excited configurations and Y number of doubly excited configurations, N is the total number of orbitals, including N$_{\textrm{o}}$ number of occupied orbitals and N$_{\textrm{v}}$ number of virtual orbitals
      \item[**] Singlet spin-adapted configuration: $\ket{S_i^b}=\frac{1}{\sqrt{2}}\left(\ket{\Phi_i^b}+\ket{\Phi_{\overline{i}}^{\overline{b}}}\right)$. When $i \neq h$, we set ${}^1\!\bigket{\Phi_{ih}^{ll}}=\frac{1}{\sqrt{2}}\left(\ket{\Phi_{i\overline{h}}^{l\overline{l}}}+\ket{\Phi_{\overline{i}h}^{\overline{l}l}}\right)$. When $i=h$, we set ${}^1\!\bigket{\Phi_{ih}^{ll}}=\bigket{\Phi_{h\overline{h}}^{l\overline{l}}}$.
      \end{tablenotes}
      
    \end{threeparttable}
    \label{tbl:selective CI}
\end{table*}
\newpage

\section{Results\cite{Results}}

\label{results}

\subsection{Impurity Population $n_1(\epsilon_d)$ and $n_2(\epsilon_d)$}

\begin{figure}[h]
  \vspace{-140pt}
\begin{subfigure}{.5\textwidth}
  \hspace*{-20mm}
  \includegraphics[width=1.5\linewidth]{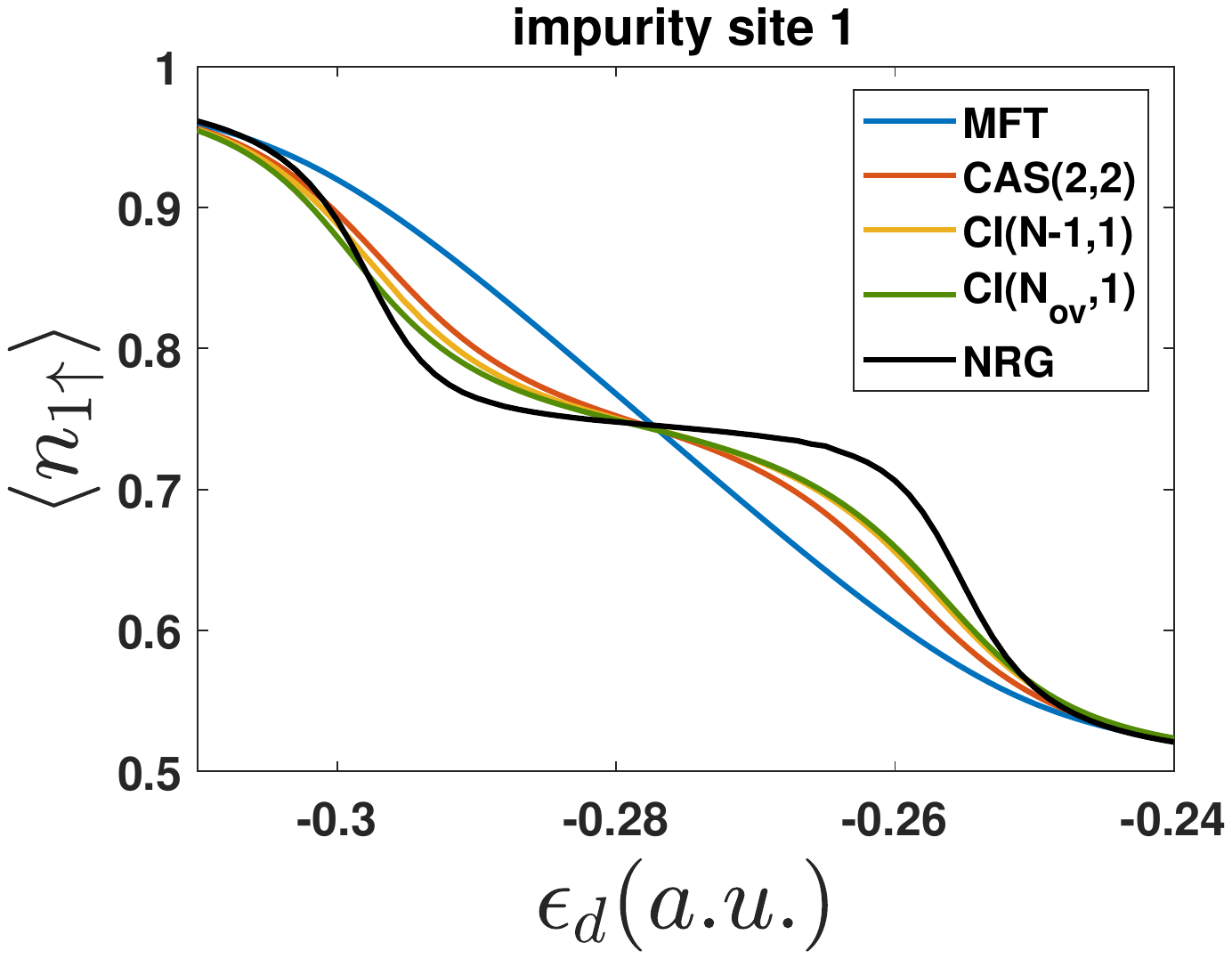}
  \vspace*{-50mm}\caption{adding more singly excited configurations}
  \label{fig:n_imp1_CIS1D}
\end{subfigure}%
\begin{subfigure}{.5\textwidth}
  \centering
  \hspace*{-20mm}\includegraphics[width=1.5\linewidth]{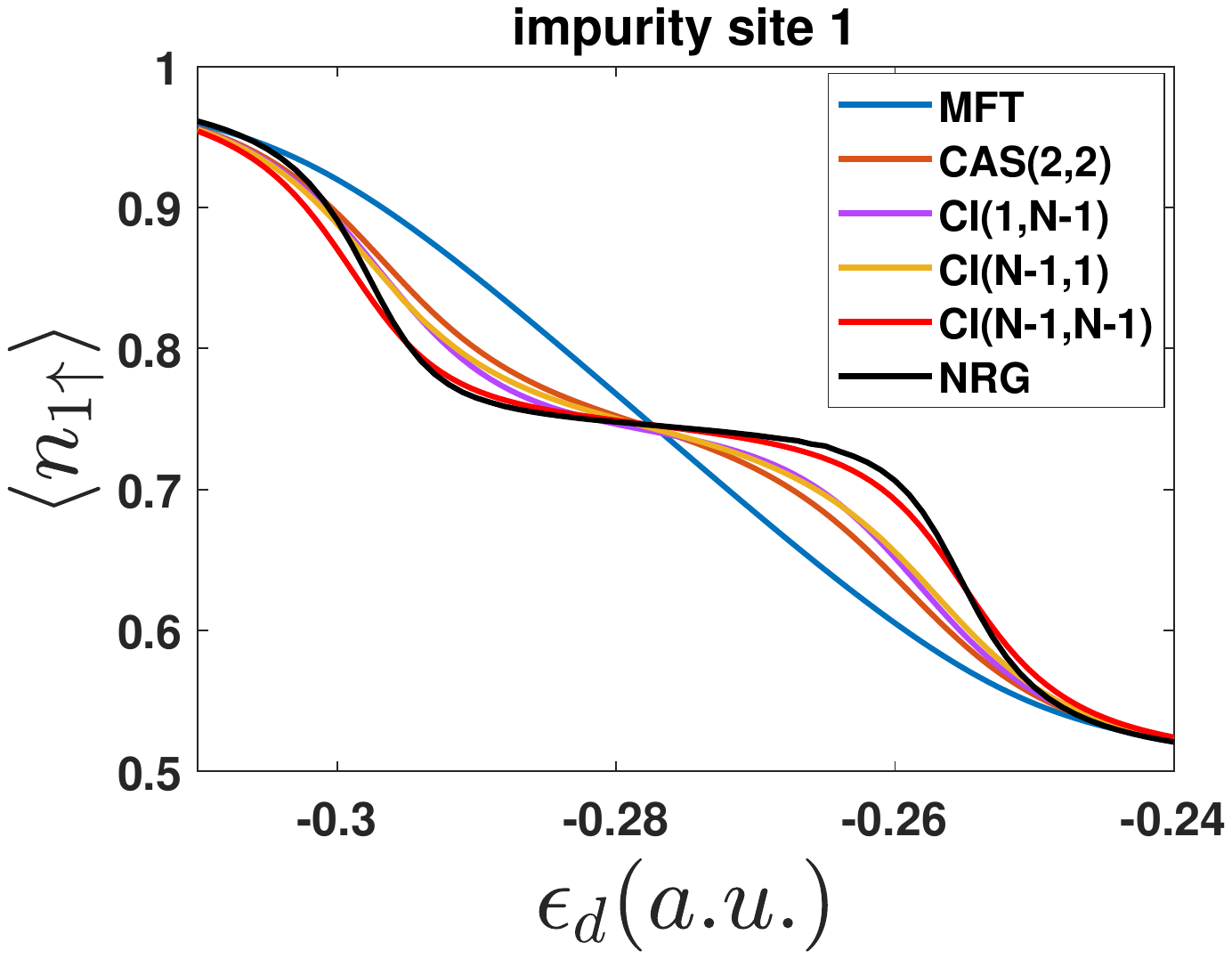}
  \vspace*{-50mm}\caption{adding more doubly excited configurations}
  \label{fig:n_imp1_CISND}
\end{subfigure}
\caption{Impurity population results from different choices of configurations for $t_d=0.2$ and $\Delta \epsilon_d=0$. (a) We include more singly excited configurations and (b) We include more doubly excited configurations. Note that adding more doubly excited configurations is more effective at recovering the plateau around $\epsilon_d=-0.29$ and $\epsilon_d=-0.26$ than is adding more singly excited configurations.}
\label{fig:n_imp_CISD}
\end{figure}

We begin by analyzing population results for the impurity, $\langle n_{1\uparrow} \rangle \equiv \langle \Psi_0|\hat{d}_{1\uparrow}^{\dagger}\hat{d}_{1\uparrow}|\Psi_0 \rangle$, according to the different selective CI methods in Table \ref{tbl:selective CI}. As a test of each method, we set $\Delta \epsilon_d = 0$, and in Fig. \ref{fig:n_imp_CISD}, we report $\langle n_{1\uparrow} \rangle$ as a function of $\epsilon_d$, the energy of the $d_1$ and $d_2$ impurities.  Since both impurities are given the same energy, we find that their populations are almost identical (not shown).  As a practical matter, in Fig \ref{fig:n_imp_CISD}, we find three different plateau regimes: 
\begin{itemize}
\item  In the range $\epsilon_d<-0.3$, $n(\epsilon_d)=4$, there are four electrons in total on the impurities and
$d_{1\uparrow},d_{1\downarrow}, d_{2\uparrow}, d_{2\downarrow}$ are all occupied.
\item  In the range $-0.3<\epsilon_d<-0.26$, $n(\epsilon_d)=3$, there are three electrons in total on the impurities. 
\item  In the range $-0.26<\epsilon_d<0.15$ (not shown completely), $n(\epsilon_d)=2$, there are two electrons in total on the impurities.
\end{itemize}

Although Fig. \ref{fig:n_imp_CISD} is limited to the region $\epsilon_d < -0.24$, two more plateaus can be identified (not shown):
\begin{itemize}
\item  In the range $0.15<\epsilon_d<0.19$, $n(\epsilon_d)=1$, there is one electron in total on the impurities.
\item  In the range $\epsilon_d>0.19$, $n(\epsilon_d)=0$, there is no electron on the impurities.
\end{itemize}
Altogether, by changing $\epsilon_d$, we can isolate four different electron transfer (ET) processes.
The first ET process happens around $\epsilon_d=-0.3$ and the second ET process happens around $\epsilon_d=-0.26$. 
 
Now, when analyzing Fig. \ref{fig:n_imp1_CIS1D}, the first thing one notices is that MFT (incorrectly) does {\em not} predict a plateau over the regime $-0.29<\epsilon_d<-0.26$, where $n(\epsilon_d) \approx 3$. Given that failure, in Fig. \ref{fig:n_imp1_CIS1D}, we consider the effect of adding in more singly excited  configurations, analyzing (in order) CAS(2,2), CI(N-1,1) and CI(N$_{\textrm{ov}}$,1).
These three methods
differ in terms of the number of singly excited configurations, but they all include exactly
one doubly excited configuration $\ket{\Phi_{h\overline{h}}^{l\overline{l}}}$. We find that, compared to the MFT results, CAS(2,2) gives a huge correction but adding more singly excited configurations does
not yield results that are significantly closer to the exact NRG results. 

Next, in Fig. \ref{fig:n_imp1_CISND}, we consider the effect of adding in more doubly excited configurations, analyzing (in order) \{ CAS(2,2), CI(N-1,1) \} versus \{ CI(1,N-1), CI(N-1,N-1)\}.  The former set includes  only one doubly excited configuration $\{\ket{\Phi_{h\overline{h}}^{l\overline{l}}}\}$  whereas the latter set includes N-1 doubly excited configurations of the form $\{{}^1\!\bigket{\Phi_{ih}^{ll}}, {}^1\!\bigket{\Phi_{hh}^{al}}\}$. Within each of these sets, we include a different number of singly excited configurations, either just $\{\ket{S_h^l}\}$  or   $\{\ket{S_i^l}, \ket{S_h^a}\}$.
Among all of these different selective CI methods, only CI(N-1,N-1) nearly matches the NRG results, and luckily with a relatively small number of configurations.

\begin{figure}[h]
  \centering
  \includegraphics[width=1\linewidth]{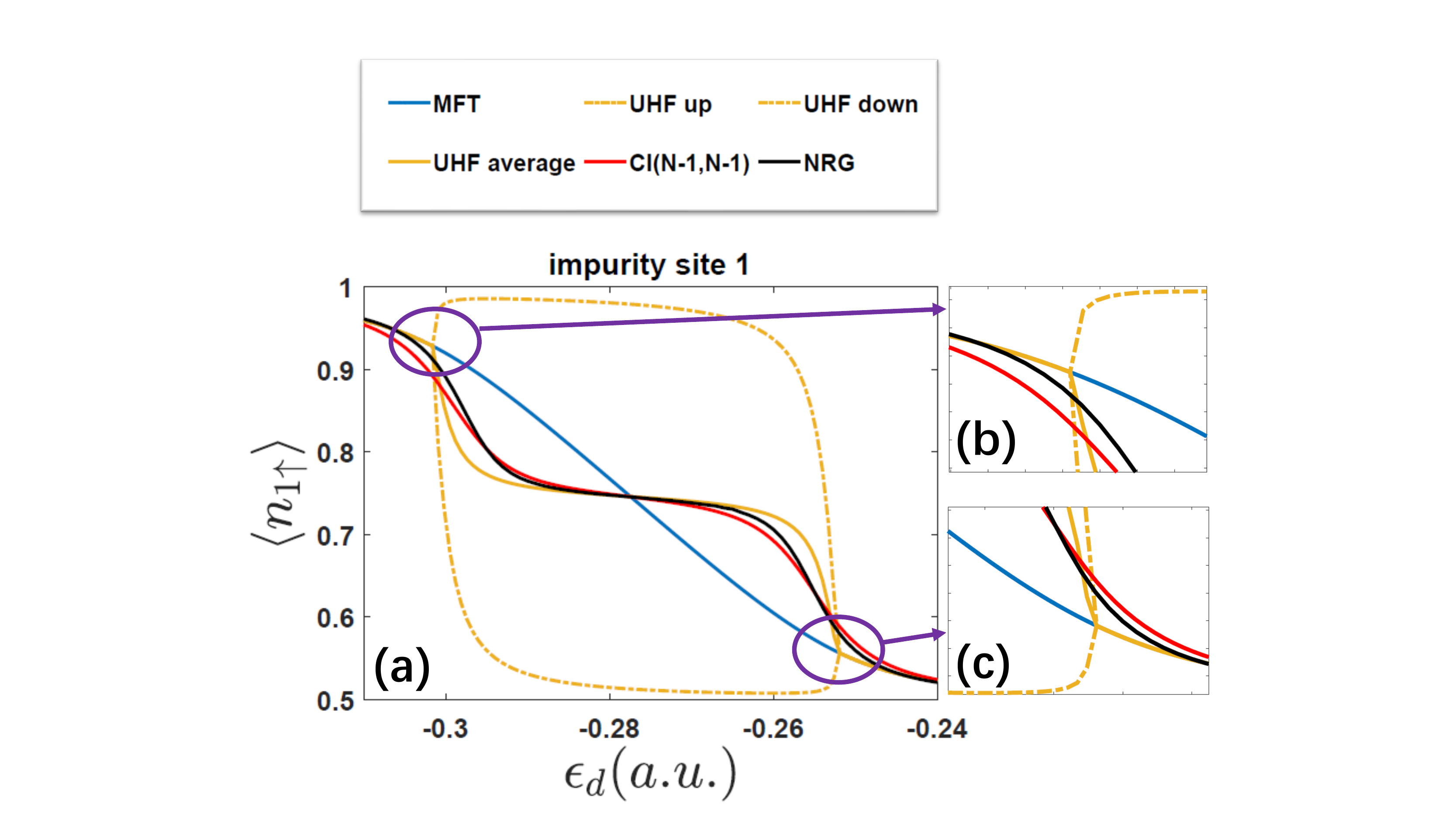}
\caption{Electron population on the impurity site 1 for $t_d=0.2$ and $\Delta \epsilon_d=0$. (a) The range is $-0.31<\epsilon_d<-0.24$, where the total number of electrons on impurities $n(\epsilon_d)$ satisfies: $4 \geq n(\epsilon_d)\geq 2$; (b,c) Zoom in of the two UHF discontinuity regions. We plot MFT (mean-field), UHF up/down (unrestricted Hartree-Fock for spin up/down electron), UHF average (averaged results of UHF up and down), CI(N-1,N-1) and NRG (the numerical renormalization group theory). NRG results are effectively exact. Note that MFT is smooth but inaccurate, whereas UHF is accurate in the plateau region but discontinuous around $\epsilon_d=-0.3$ and $\epsilon_d=-0.25$.}
\label{fig:n_imp}
\end{figure}

At this point, having analyzed quite a few restricted CI approaches, in 
 Fig. \ref{fig:n_imp}, we compare the most promising restricted CI method [CI(N-1,N-1)]
against the simplest  unrestricted method, unrestricted Hartree Fock (UHF).
For UHF, one breaks symmetry such that $\langle n_{1\uparrow} \rangle \ne \langle n_{1\downarrow} \rangle$.  For this reason, in order to compare
UHF results vs. NRG results, we will need to average the two solutions:
\begin{eqnarray}
  \bar{n}_1 = \frac{\langle n_{1\uparrow} \rangle + \langle n_{1\downarrow} \rangle}{2}
  \label{eq:n1average}
\end{eqnarray}
 From Fig. \ref{fig:n_imp}, one can clearly see that the UHF average in Eq. \ref{eq:n1average} reproduces the plateau 
region where $n(\epsilon_d)$ = 3 very well. Nevertheless, as the inserts Figs. \ref{fig:n_imp}(b,c) show clearly,
the UHF ansatz  introduces an artificial discontinuity at the edge points of the plateau region  ($\epsilon_d \approx -0.3$ or $\epsilon_d \approx -0.25$) where there is a Coulson-Fisher point \cite{coulson1949xxxiv} and the solution switches between restricted and unrestricted wavefunctions (and electron transfer occurs).
At these points, $\frac{\partial n(\epsilon_d)}{\partial \epsilon_d}$ is clearly discontinuous. For this reason, given our long term interest in dynamics, below we will not focus too much on UHF solutions (though see also Sec. \ref{discussion2}. 1).

\newpage

\subsection{Total Energy}
\begin{figure}[htbp]
\captionsetup[subfigure]{labelformat=empty}
\centering
\vspace*{-70mm}
\begin{subfigure}[t]{1\textwidth}
    \centering
    \includegraphics[width=1\linewidth]{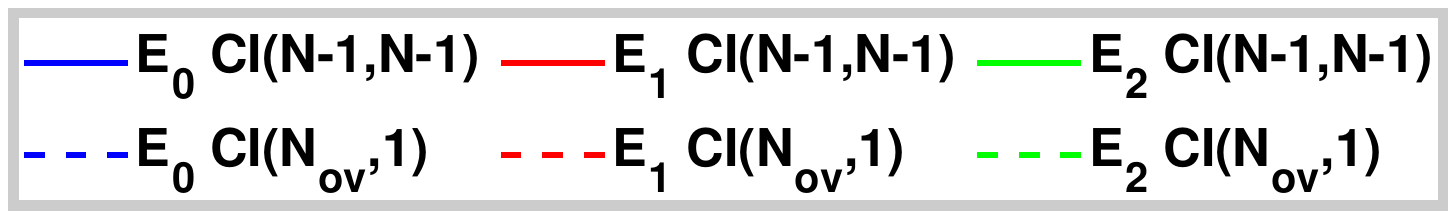}
\end{subfigure}%

\medskip

\begin{subfigure}[t]{.5\textwidth}
  \centering
  \vspace*{-150mm}
  \hspace*{-20mm}\includegraphics[width=1.5\linewidth]{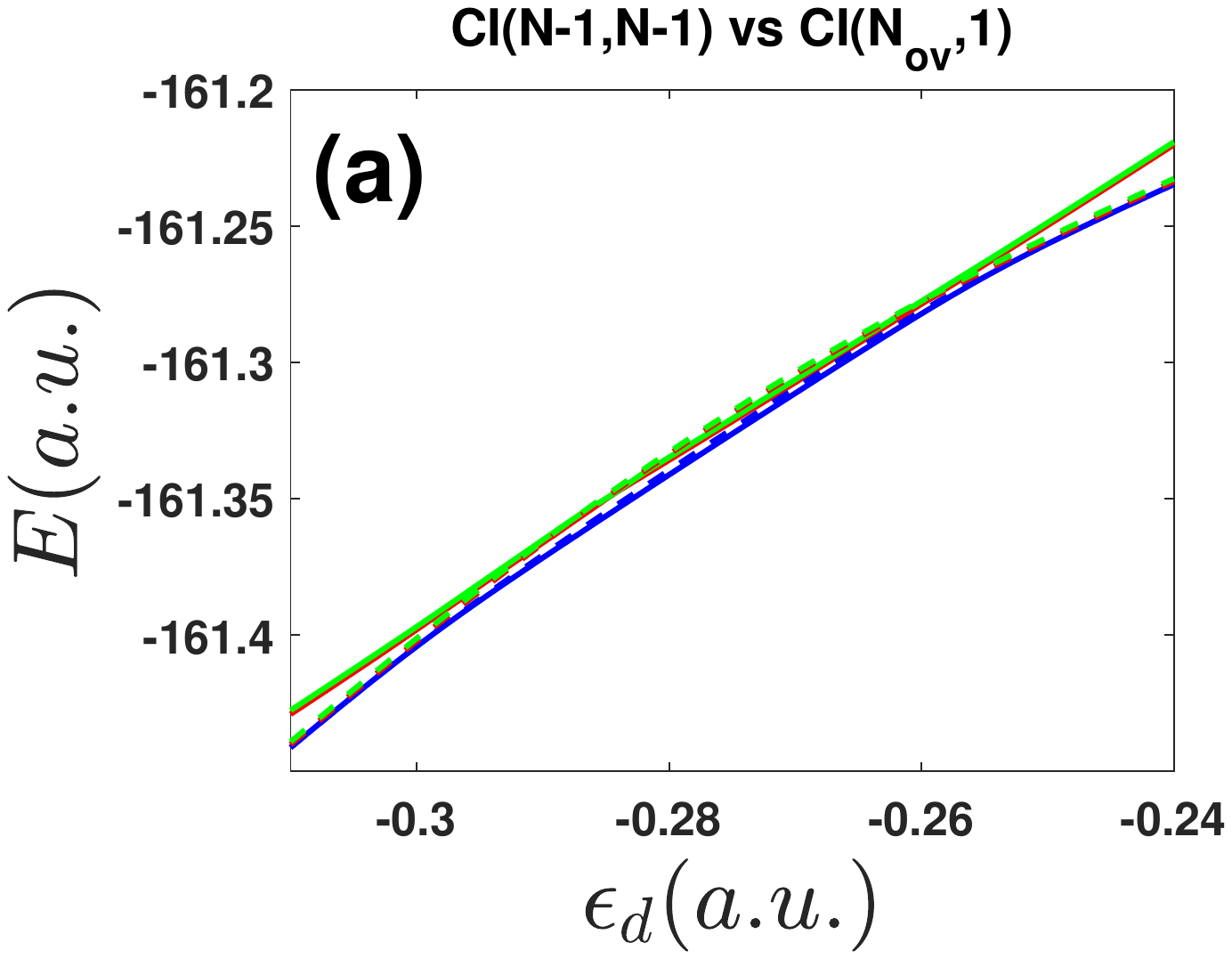}
  \vspace*{-35mm}\caption{}
  \label{fig:Energy_4elec}
\end{subfigure}%
\begin{subfigure}[t]{.5\textwidth}
  \centering
  \vspace*{-150mm}
  \hspace*{-20mm}\includegraphics[width=1.5\linewidth]{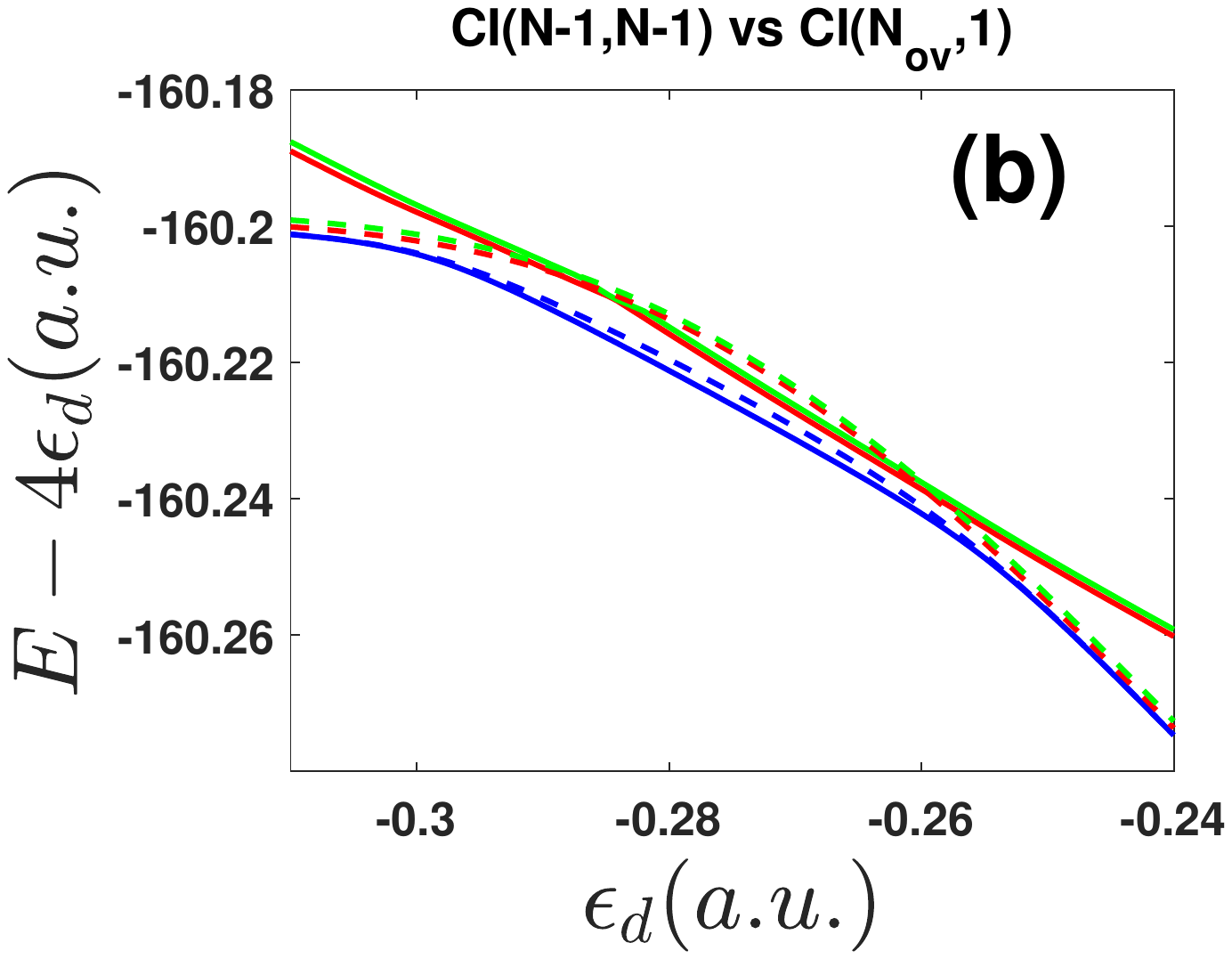}
  \vspace*{-35mm}\caption{}
  \label{fig:E-4ed}
\end{subfigure}

\vspace*{-25mm}
\caption{CI(N-1,N-1) (solid line) and CI(N$_\textrm{{ov}}$,1) (dashed line) results for the lowest three state energies (eigenvalues) for $t_d=0.2$ and $\Delta \epsilon_d=0$. Within the range $-0.31<\epsilon_d<-0.24$, the total number of electrons on the impurities is between 4 and 2 ($2\leq n(\epsilon_d) \leq 4$). (a) Total energy $E$ versus $\epsilon_d$. (b) $E-4\epsilon_d$ versus $\epsilon_d$; we plot this quantity in order to compare CI(N-1,N-1) results versus CI(N$_{\textrm{ov}}$,1) results more clearly near the crossing point. Note that in Fig. \ref{fig:E-4ed}, CI(N-1,N-1) finds the lower ground state and excited states energies when $-0.3<\epsilon_d<-0.26$.}
\label{fig:Energy_smallcisnd_cis1d}
\end{figure}

Beyond impurity populations, if one wants to either calculate thermodynamic quantities or simulate dynamical trajectories, the most important quantity of interest is the total energy of the universe (molecule + metal).  To best understand the merits
of the CI approaches described above, in Fig. \ref{fig:Energy_smallcisnd_cis1d}, we plot
the first three state energies (or Hamiltonian eigenvalues) as  calculated by \cite{NRGenergy}
\begin{itemize}
\item CI(N-1,N-1), the method which performed best above at recovering impurity populations.
\item CI(N$_{\textrm{ov}}$,1), the CI method with the largest number of  configurations; see Table \ref{tbl:selective CI} 
\end{itemize}
In Fig. \ref{fig:E-4ed}, 
we find that in the energy regime $-0.3<\epsilon_d<-0.26$, where $n(\epsilon_d)=3$, including doubly excited configurations is crucial as far as minimizing the ground state energy. And including doubly excited configurations is more important than including singly excited configurations, which seemingly agrees with the Brillouin's theorem ($\bracket{\Phi_{\textrm{HF}}}{S_i^a}=0$). This finding helps explain why
    the CI(N-1,N-1) method performed so well at recovering $\langle n_{1\uparrow} \rangle= \langle \Psi_0 | d_{1\uparrow}^{\dagger}d_{1\uparrow} | \Psi_0 \rangle$ in Fig \ref{fig:n_imp}.
However, note that (in fairness), Fig. \ref{fig:E-4ed} also makes clear that, when calculating excited states (especially in the $n(\epsilon_d)=4$ and $n(\epsilon_d)=2$ regions), CI(N$_{\textrm{ov}}$,1)
finds significantly lower variational energies \cite{CIS1D}.

\begin{figure}[h]
    \centering
    \vspace*{-35mm}
    \hspace*{-20mm}\includegraphics[width=1.3\linewidth]{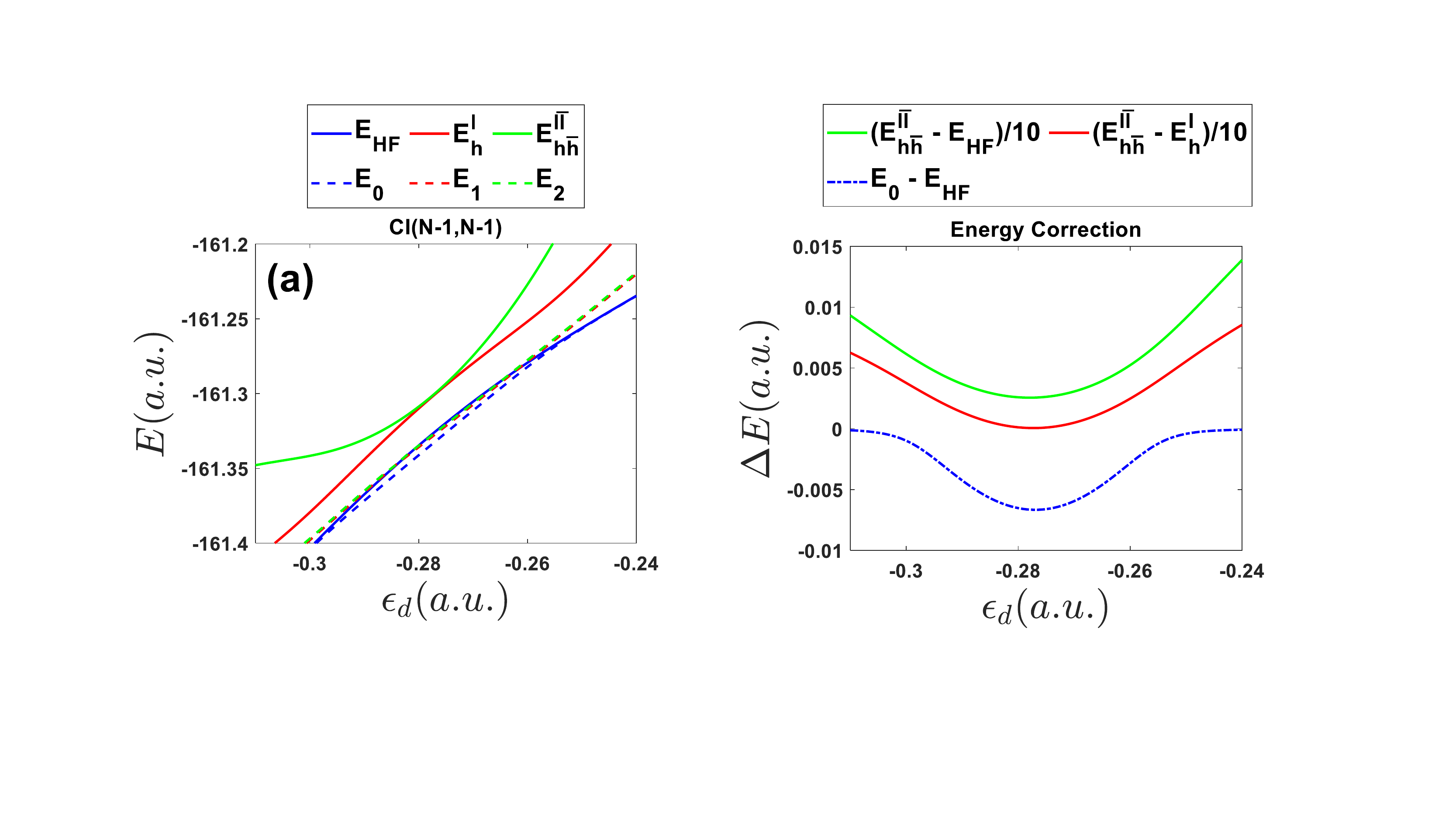}


\vspace*{-30mm}\caption{ 
(a) The three lowest raw configuration energies (solid line) are labeled as E$_{\textrm{HF}}$, E$_\textrm{h}^\textrm{l}$ and E$_{\textrm{h}\overline{\textrm{h}}}^{\textrm{l}\overline{\textrm{l}}}$, corresponding to the energies of the Hartree-Fock ground state configuration, the HOMO-to-LUMO singly excited configuration and the HOMO-to-LUMO doubly excited configuration, respectively. The three lowest CI(N-1,N-1) eigenvalues (dashed line) are labeled as E$_0$, E$_1$ and E$_2$. (b) The relationship between the ground state energy correction (blue dash-dotted line) and the energy differences between configurations (red and green solid line). 
Fig. \ref{fig:Energy}b demonstrates that the absolute value of the ground state energy correction is maximum when the energy of the HOMO-to-LUMO doubly excited configuration is closest both to the energy of the HF ground state configuration and to the energy of the the HOMO-to-LUMO singly excited configuration. Note that, with parameter settings $t_d=0.2$ and $\Delta \epsilon_d=0$, the maximum CI(N-1,N-1) correction to the ground state can be as large as 0.005 hartrees.}
\label{fig:Energy}
\end{figure}

Finally, having convinced ourselves of the importance of adding in doubly excited configurations, it is instructive to compare the energies of the final CI(N-1,N-1) eigenvalues
with the set of essential CAS(2,2) configurations, $\left\{ \bigket{\Phi_{\textrm{HF}}}, \bigket{S_h^l}, \bigket{\Phi_{h\overline{h}}^{l\overline{l}}}\right\}$; such a comparison will hopefully yield simple insight as to if/when a sophisticated CI approach is needed.  In Fig. \ref{fig:Energy}(a), we plot the energies of these essential configurations, as well as the three lowest energies found after the CI(N-1,N-1) diagonalization. We find that, when the  energy of the doubly excited configuration $\bigket{\Phi_{h\overline{h}}^{l\overline{l}}}$ approaches the energy of the HF state $\bigket{\Phi_{\textrm{HF}}}$ (and nearly crosses the energy of the singly excited configuration $\bigket{S_h^l}$), there is a huge correction to the ground state energy. This avoided crossing occurs in the regime $\epsilon_d \approx -0.28$, $n(\epsilon_d) \approx 3$;  see Fig. \ref{fig:Energy}(b). 

As a side note, the reviewer can also discern from Fig. \ref{fig:Energy}  that, even for a modest CI calculation (e.g. CI(N-1,N-1)), the predicted first excited state energy, E$_1$ CI(N-1,N-1), is far away from the energy of the HOMO-LUMO transition, E$_\textrm{h}^\textrm{l}$.  Thus, as mentioned above, one must be careful in how one assesses the value of excited state calculations for a large CI calculation with a continuum of states; in this instance, excited states need to be understood as part of a dense set of states and the accuracy of these states can only be determined dynamically.



\newpage

\section{Discussion}
\label{discussion2}

\subsection{Electron Transfer from an Open Shell Impurity Singlet to the Metal}

The results presented above should convince the reader that, at least for a model Hamiltonian with two impurity sites, one can recover reasonable results using basic configuration interaction theory. Now,  one might be tempted to think that having two impurity sites is not so different from having one (dressed) impurity site \cite{jin2020configuration}. Unfortunately, the latter statement is incorrect. After all, in a certain parameter regime, one should find dynamics characterized by the Fig. \ref{fig:openshell}:

\begin{figure}
    \centering
    \includegraphics[width=0.5\linewidth]{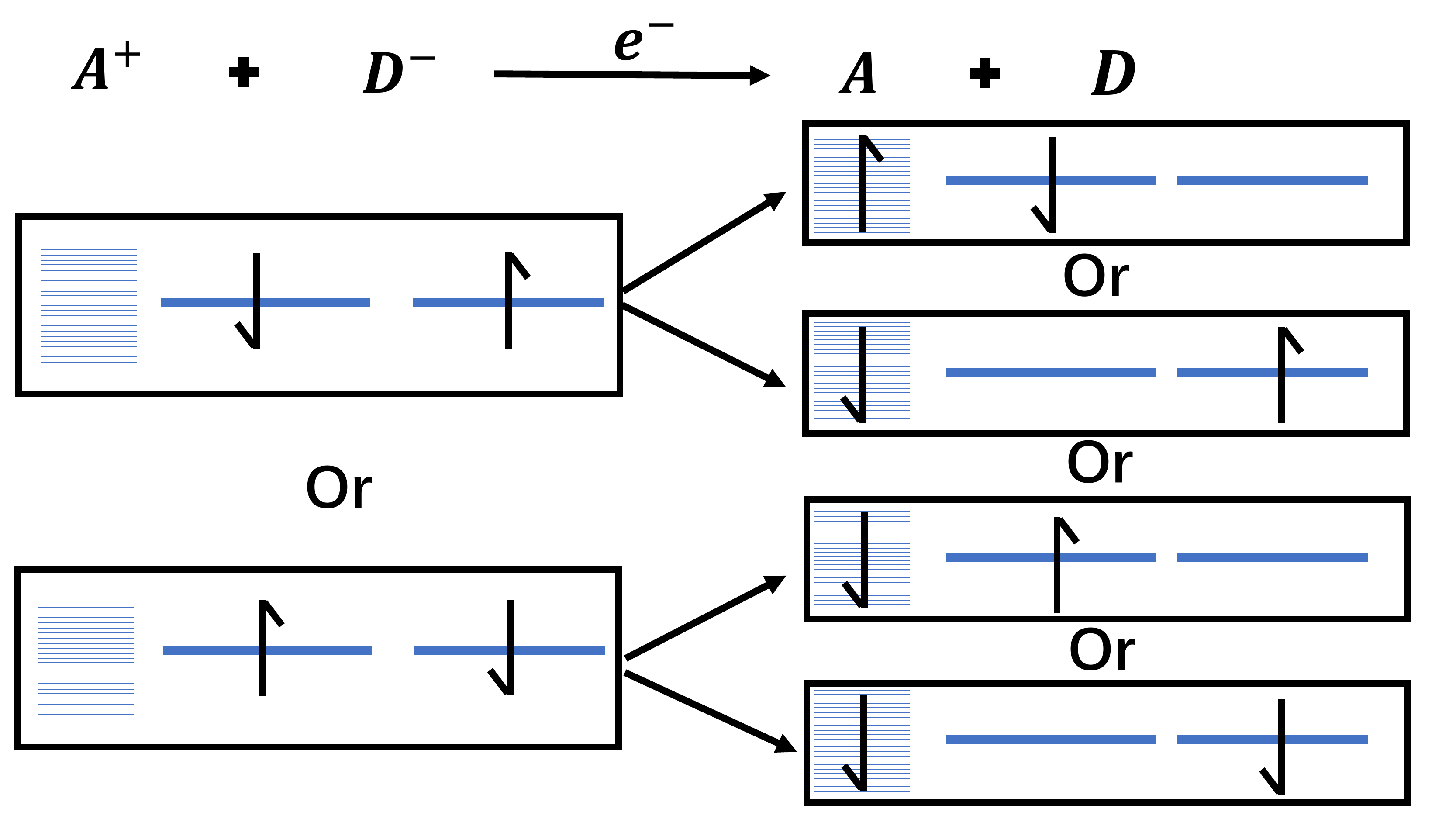}
    \caption{A schematic figure for an open-shell impurity singlet electron transfer. Note that when $t_d$ is small, the electronic state of the impurity tends to have more of an open-shell character.}
    \label{fig:openshell}
\end{figure}

In other words, one can imagine electron transfer from an {\em open shell singlet} residing on the 
impurity (with two sites!) to the metal.  Such rich physics cannot be captured by a one-site impurity model. And yet, capturing such physics would clearly be essential for modeling how a chemical 
bond breaks on a metal surface.

\label{parameter test}
\subsubsection{Vary $t_d$ (keeping $\Delta\epsilon_d=0$)}
\label{vary td}
\begin{figure}[htbp]
    \vspace*{-15mm}
\begin{subfigure}[t]{.5\textwidth}
    \centering
    \hspace*{-60mm}\includegraphics[width=2.5\linewidth]{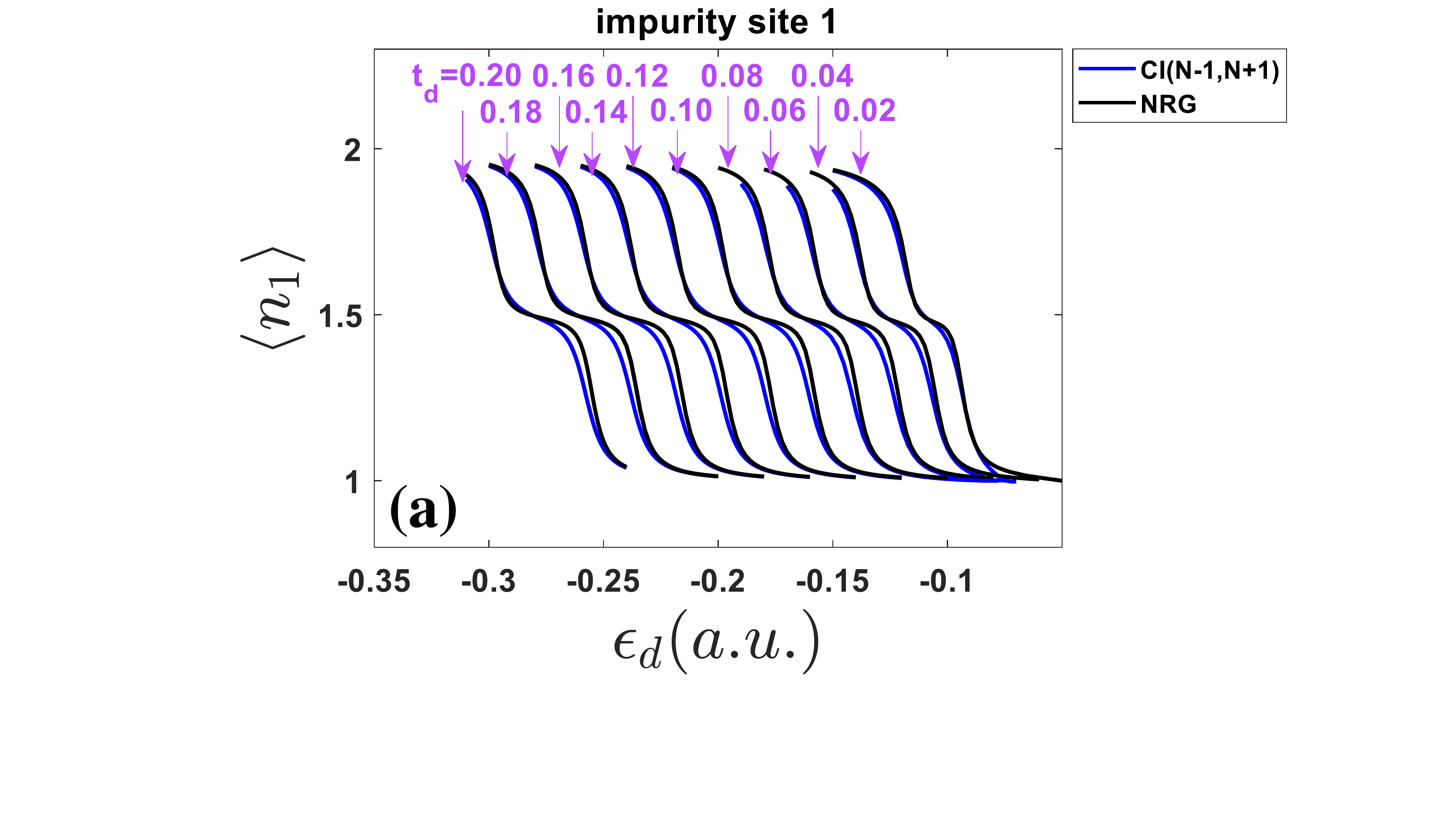}
    \label{fig:smallcisndplus}
    \end{subfigure}
    
\begin{subfigure}[t]{.5\textwidth}
    \centering
    \vspace*{-15mm}
    \hspace*{-60mm}\includegraphics[width=2.5\linewidth]{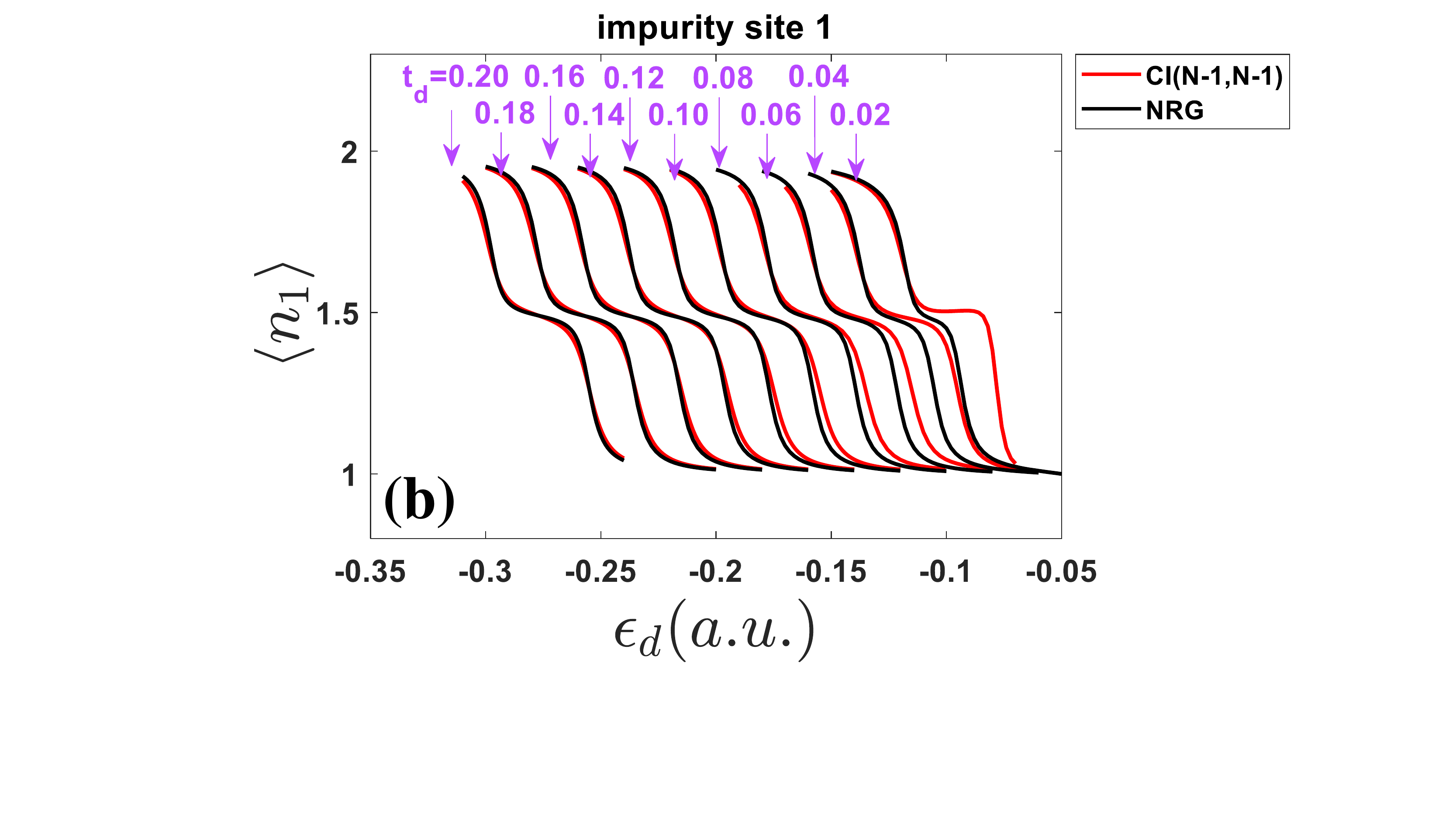}
    \label{fig:smallcisnd}
    \end{subfigure}
    \vspace*{-20mm}\caption{CI(N-1,N+1) (blue solid line in (a)), CI(N-1,N-1) (red solid line in (b)) and NRG (black solid line) results for impurity population as a function of impurity energy $\epsilon_d$ for different $t_d$ with $\Delta \epsilon_d=0$. The hopping strength between two impurity sites $t_d$ ranges from $0.02$ to $0.2$. Note that CI(N-1,N-1) fails for small $t_d$ while CI(N-1,N+1) results match the NRG results for all values of $t_d$.}
    \label{fig:n_imp1_td}
\end{figure}

To better understand the nature of electron transfer from an {\em open shell singlet} on an impurity into a metal substrate, 
we have rerun all the calculations presented above with different parameters for $t_d$. After all, when $t_d$ is large, we can expect large
hybridization of the impurity orbitals (and therefore, if the number of electrons is even, the impurity should prefer to be in a closed shell singlet). However, if $t_d$ is small and the number of electrons is even, we can expect the impurity will prefer an {\em open shell singlet} configuration (as in Fig. \ref{fig:openshell}). This is the same as a Mott transition. \cite{mott1968metal}

In Fig. \ref{fig:n_imp1_td}b, we benchmark CI(N-1,N-1) for several different values of  $t_d$, ranging from $t_d=0.2$ to $t_d=0.02$. For large values of $t_d$ ($t_d>0.1$), we find that indeed, CI(N-1,N-1) results do match the exact NRG results. However, for small
values of $t_d$ ($t_d<0.1$), CI(N-1,N-1) fails. In Fig. \ref{fig:n_imp1_td}b, we find an anomalously large (and incorrect) plateau when $t_d=0.02$.

In order to address this failure, one can argue that it is appropriate to include one more doubly excited configuration. 
The reason is as follows.
Consider the case when $\langle n \rangle/2 =\langle n_1 \rangle = \langle n_2 \rangle = 1.5$ ($\langle n_2 \rangle$ not shown in the figure) in Fig. \ref{fig:n_imp1_td}b. For this value of $\epsilon_d$, if we diagonalize the mean-field impurity Hamiltonian $\hat{H}_{imp}$,
\begin{equation}
    \hat{H}_{imp}=
    \begin{pmatrix}
        \epsilon_d+U \langle n_2 \rangle & t_d \\
        t_d & \epsilon_d+U \langle n_1 \rangle
        \end{pmatrix}
\end{equation}
we recover two orbital energies (two eigenvalues):
\begin{equation}
\begin{aligned}
    \epsilon_{h-1}=\epsilon_d+U \langle n \rangle/2 - t_d \\
    \epsilon_h=\epsilon_d+U \langle n \rangle/2 + t_d
    \end{aligned}
\end{equation}
Thereafter we must place 3 electrons into these two orbitals since $\langle n \rangle = 3$. Now, because of electron-electron repulsion, one can expect
that a lower energy ground state can be found by unrestricting the calculation,
leading to a new set of energies for which the alpha orbitals are shifted in energy from the beta orbitals (by an amount that we call $U_{\textrm{eff}}$).
In such a case, the HOMO and HOMO-1 orbitals will be different for alpha and beta spin (without loss of generality, we assume $\epsilon_{i} < \epsilon_{\overline{i}}$):
\begin{equation}
    \begin{aligned}
            \epsilon_{h-1}&=\epsilon_d+U \langle n \rangle/2 - t_d \\
            \epsilon_{\overline{h-1}}&=\epsilon_d+U \langle n \rangle/2 - t_d + U_{\textrm{eff}}\\
            \epsilon_{h}&=\epsilon_d+U \langle n \rangle/2 + t_d \\
            \epsilon_{\overline{h}}&=\epsilon_d+U \langle n \rangle/2 + t_d +U_{\textrm{eff}}\\
    \end{aligned}
\end{equation}
Now if $t_d$ is large in the sense that $2t_d > U_{\textrm{eff}}$ (for our results, $t_d=0.2$, $U=0.1$, so we assume $U_{\textrm{eff}}$ is smaller than 0.1), the energy ordering of the orbitals is standard: $\epsilon_{h-1} < \epsilon_{\overline{h-1}} < \epsilon_{h} < \epsilon_{\overline{h}}$. However, if $t_d$ is small in the sense that $2t_d < U_{\textrm{eff}}$ (e.g. $t_d=0.02$), the energy ordering of the orbitals inverts: $\epsilon_{h-1} <  \epsilon_{h} < \epsilon_{\overline{h-1}} < \epsilon_{\overline{h}}$. Thus, in the small $t_d$ limit, if we consider the case when two electrons are excited from occupied orbitals to virtual orbitals, the doubly excited configurations $\ket{\Phi_{h-1\overline{h-1}}^{l\overline{l}}}$ as well as $\ket{\Phi_{h\overline{h}}^{l+1\overline{l+1}}}$ should play an important role in a CI calculation.
For this reason, we have included one more CI method in Table \ref{tbl:selective CI}, namely CI(N-1,N+1), for which we include two extra configurations, $\ket{\Phi_{h-1\overline{h-1}}^{l\overline{l}}}$ and $\ket{\Phi_{h\overline{h}}^{l+1\overline{l+1}}}$. 
In Fig. \ref{fig:n_imp1_td}a, we demonstrate that CI(N-1,N+1) does recover the correct populations quantitatively.

\begin{figure}[htbp]
    \centering
    \vspace*{-22mm}
    \hspace*{-110mm}\includegraphics[width=2.2\linewidth]{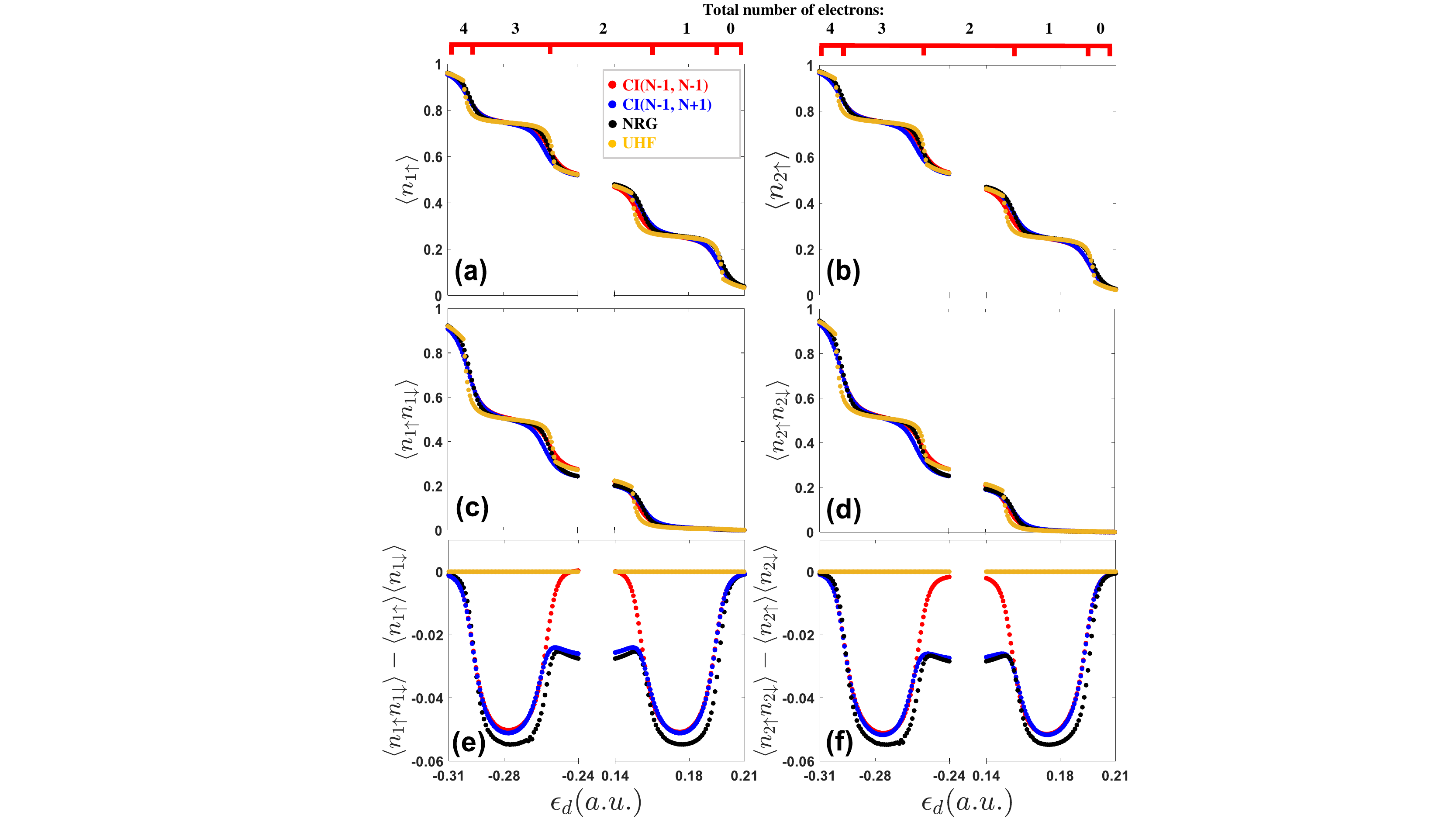}
    \caption{Single occupancy, double occupancy and occupancy correlation on each impurity site for $t_d = 0.2$ and $\Delta \epsilon_d=0$. (a) Single occupancy on the impurity site 1; (b) Single occupancy on the impurity site 2; (c) Double occupancy on the impurity site 1; (d) Double occupancy on the impurity site 2; (e) Correlation on the impurity site 1 ($\langle n_{1\uparrow}n_{1\downarrow} \rangle - \langle n_{1\uparrow} \rangle \langle n_{1\downarrow} \rangle $); (f) Correlation on the impurity site 2 ($\langle n_{2\uparrow}n_{2\downarrow} \rangle - \langle n_{2\uparrow} \rangle \langle n_{2\downarrow} \rangle $). Note that CI(N-1,N-1) performs well as compared with NRG. However, also note that there is not very much electron-electron correlation in Figs. \ref{fig:n_td02}(e-f) (the maximum is only 0.06). }
    \label{fig:n_td02}
\end{figure}

\begin{figure}[htbp]
    \centering
    \vspace*{-22mm}
    \hspace*{-100mm}\includegraphics[width=2.2\linewidth]{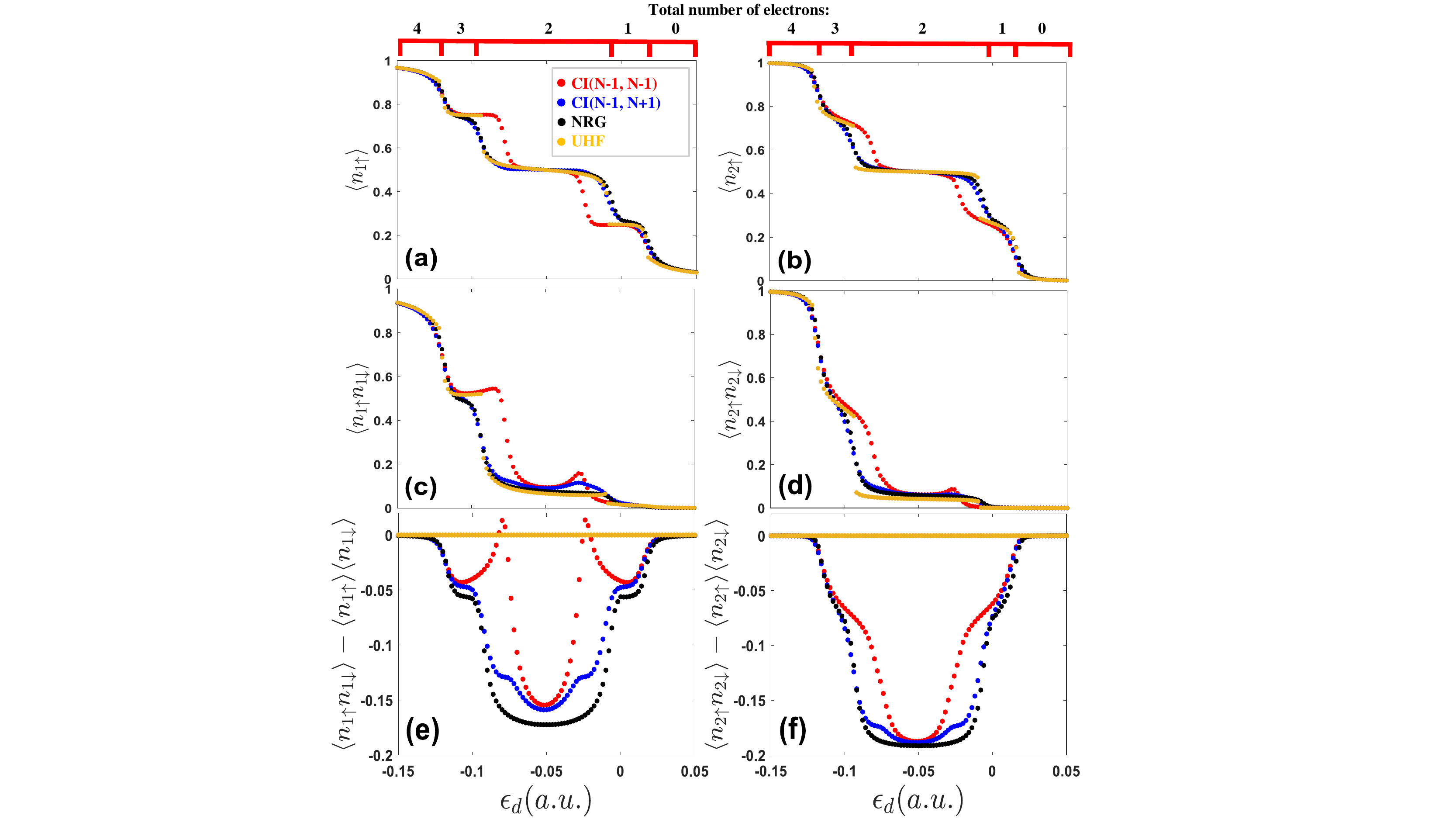}
    \caption{Single occupancy, double occupancy and occupancy correlation on each impurity site for $t_d = 0.02$ and $\Delta \epsilon_d=0$. (a) Single occupancy on the impurity site 1; (b) Single occupancy on the impurity site 2; (c) Double occupancy on the impurity site 1; (d) Double occupancy on the impurity site 2; (e) Correlation on the impurity site 1 ($\langle n_{1\uparrow}n_{1\downarrow} \rangle - \langle n_{1\uparrow} \rangle \langle n_{1\downarrow} \rangle $); (f) Correlation on the impurity site 2 ($\langle n_{2\uparrow}n_{2\downarrow} \rangle - \langle n_{2\uparrow} \rangle \langle n_{2\downarrow} \rangle $). Note that in Figs. (e-f), the depth of the correlation single well is about 0.2, which is three times bigger than that for $t_d=0.2$ case (which was plotted in Fig. \ref{fig:n_td02}). For this data set, CI(N-1,N+1) vastly outperforms CI(N-1,N-1).}
    \label{fig:n_td002}
\end{figure}

Now, the argument above may appear cyclical and flawed. After all, the interpretation above was entirely predicated
on the idea that, for small $t_d$, one would find an {\em open shell singlet} on the impurity -- and yet we never actually proved as much.
To verify that, indeed, an {\em open shell singlet} appears, in Figs. \ref{fig:n_td02} and \ref{fig:n_td002}, we analyze (in detail) the electronic structure
of the impurity sites across the whole $\epsilon_d$ range for the different $t_d$ values, one large ($t_d=0.2$) and one small ($t_d=0.02$).
In Figs. \ref{fig:n_td02} and \ref{fig:n_td002}, we plot single occupancy results (a-b) ($\langle n_{1\uparrow} \rangle$ and $\langle n_{2 \uparrow} \rangle$), double occupancy results (c-d) and the correlation between single and double occupancy (e-f). We plot results for CI(N-1, N-1), CI(N-1, N+1) and UHF (and all relative to exact NRG calculations).

For $t_d=0.2$, in Fig. \ref{fig:n_td02}, we find that, as far at the total number of electrons present (in Figs. \ref{fig:n_td02}(a-d)), the CI(N-1,N-1) and CI(N-1,N+1) results are  nearly identical, and they nearly agree with the exact NRG results (as does UHF).  Now if one looks closely in the regions
$\epsilon_d \in \left[-0.25, -0.24\right]$ and $\epsilon_d \in \left[0.14, 0.15\right]$, there are small differences. 
Indeed, in these two $\epsilon_d$ regions, where the total number of electrons in the molecule is changing from 3 to 2 and from 2 to 1, respectively,
the plot of correlation (in Figs. \ref{fig:n_td02}(e-f)) makes clear that CI(N-1,N+1) and CI(N-1,N-1) are not identical; and over the entire region where $n(\epsilon_d) =2$, i.e. 
$\epsilon_d \in \left[-0.24, 0.14\right]$,  CI(N-1,N+1) agrees with NRG (whereas CI(N-1,N-1) does not).  Nevertheless, one should
note that the scale on Figs. \ref{fig:n_td02}(e-f) is not very large (as compared with Figs. \ref{fig:n_td002}(e-f)). One should also note that,
in this figure,  the correlation between single and double occupancy is not maximized in the central $n(\epsilon_d) = 2$ region (where $\epsilon_d \in \left[-0.24, 0.14\right]$), 
but rather in the outer $n(\epsilon_d) = 3$ ($\epsilon_d=-0.28$)  and $n(\epsilon_d) = 1$ ($\epsilon_d=0.18$) regions. Altogether, this data suggests that electron
correlation exists (but is not very strong) for $t_d=0.2$, which explains why
CI(N-1,N-1) performs so well in Figs. \ref{fig:n_td02}(a-d).

Next, let us turn to  Fig. \ref{fig:n_td002}, where we plot results for the case $t_d=0.02$.
Here, we immediately see enormous differences between CI(N-1,N-1) and the exact NRG results both in terms
of the single and double occupancy results. At certain values of $\epsilon_d$, UHF can nearly match the NRG results, but not always,
especially in the regions of electron transfer (where clear discontinuities arise at each step of the curve). By contrast to the other 
methods, the CI(N-1,N+1) results match the NRG results quite well at almost all points. One can draw the same conclusions from Figs. \ref{fig:n_td002}(e-f)
with regards to electron correlation.  Finally, note that, in contrast to the case $t_d=0.2$, here we find the strongest correlation effects
within the middle range for $\epsilon_d$ ($\epsilon_d \approx -0.05$, $n(\epsilon_d) \approx 2$), confirming our premise that an {\em open shell singlet} is prominent for the case of small $t_d$.  Note also that the correlation strength is about three times as big for the $t_d=0.02$ case as for the $t_d=0.2$ case.
Overall, the conclusions from this data are that, if we include just two extra doubly excited configurations, we can really recover the lion's share
of electron correlation for a two-site impurity model on a metal surface.

\newpage

\subsubsection{Vary $\Delta \epsilon_d$ (keeping $t_d=0.2$)}
\begin{figure}[htbp]
    \vspace*{-15mm}
\begin{subfigure}[t]{.5\textwidth}
    \centering
    \hspace*{-60mm}\includegraphics[width=2.5\linewidth]{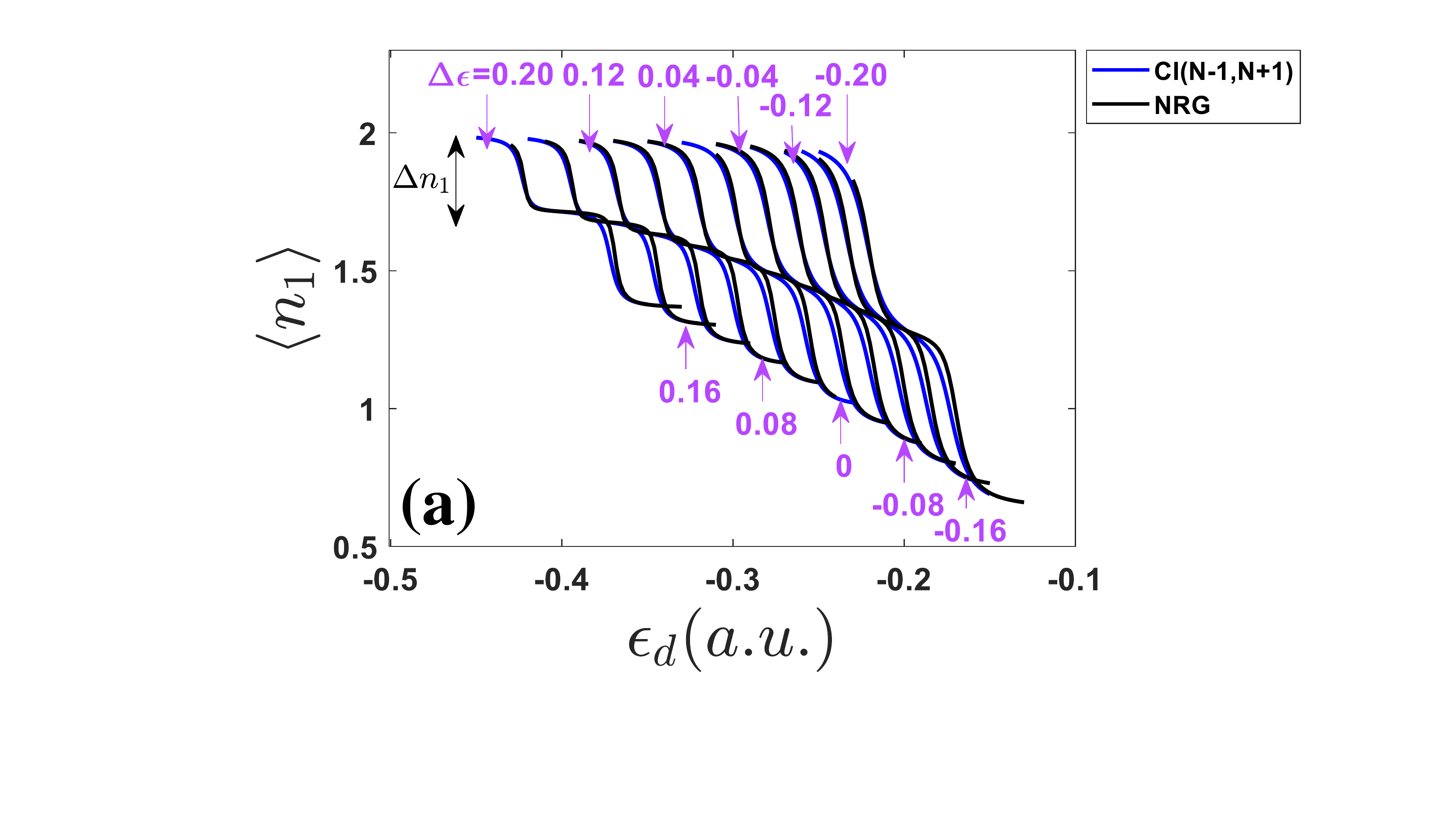}
    \end{subfigure}
    
\begin{subfigure}[t]{.5\textwidth}
    \centering
    \vspace{-55pt}
    \hspace*{-60mm}\includegraphics[width=2.5\linewidth]{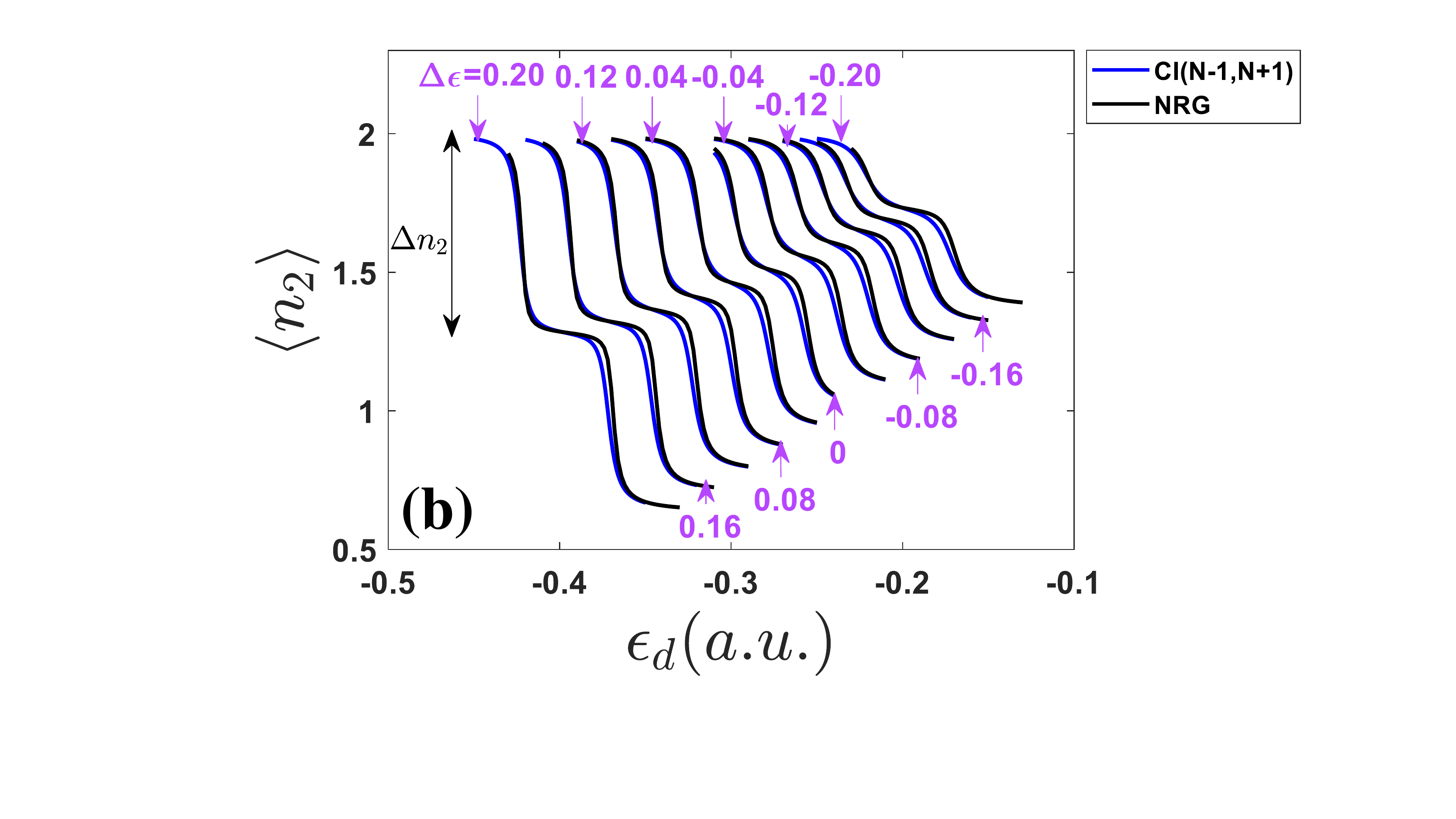}
\end{subfigure}
    \vspace*{-10mm}\caption{CI(N-1,N+1) (blue solid line) and NRG (black solid line) results for impurity population on (a) the impurity site 1 and (b) the impurity site 2 as a function of impurity energy $\epsilon_d$ for different $\Delta\epsilon_d$ with $t_d=0.2$. Here we define: $\Delta \epsilon_d \equiv \epsilon_{d2} - \epsilon_{d1}$. The relative energy difference between two impurity sites $\Delta\epsilon_d$ ranges from $-0.2$ to $0.2$. $\Delta n_1$ and $\Delta n_2$ represent the longitudinal distance of the first plateau and the second plateau for the impurity site 1 and the impurity site 2, respectively.  Note that CI(N-1,N+1) results match the NRG results for all different $\Delta\epsilon_d$.}
    \label{fig:n_imp_de}
\end{figure}

So far, within this manuscript, we have always insisted that the two sites have the same energy ($\epsilon_{d1} = \epsilon_{d2} = \epsilon_d$).  At this point, we will break this assumption as another means of testing  the quality of the CI approaches above. Let $\Delta \epsilon_d \equiv \epsilon_{d2} - \epsilon_{d1}$.
In Fig. \ref{fig:n_imp_de}, for different $\Delta \epsilon_d$ (ranging from $\Delta \epsilon_d/t_d=-1$ to $\Delta \epsilon_d/t_d=1$),  we plot population results
as a function of $\epsilon_d$ over a range so that the
total  number of electrons on the two impurities changes from 4 to 3 to 2.
Our focus here will be on the first electron transfer process (i.e. the drop between the first two plateaus farthest on the left).  Here, we see that when one electron is transferred from the molecule to the metal, this transfer occurs at different values of $\epsilon_d$ (depending on $\Delta \epsilon_d$). 
If we define $\Delta n_1, \Delta n_2$ to be the number of electrons extracted from 
the impurity site 1 and the site 2 (respectively) at the first plateau, we must obviously have $\Delta n_1 +  \Delta n_2 = 1$.  As we can see from Fig. \ref{fig:n_imp_de}, when $\Delta \epsilon_d$ decreases, $\Delta n_1$ increases and $\Delta n_2$ decreases. In other words, the impurity site with a higher ionization energy will lose more electronic density during the first electron transfer process. 

Three simple limits can be identified here: 

\begin{enumerate} 

\item When $\Delta \epsilon_d=0$, $\Delta n_1=\Delta n_2=0.5$.

\item  When $\Delta \epsilon_d \gg |t_d|$, $\Delta n_1 \to 0, \Delta n_2 \to 1$.

\item  When $\Delta \epsilon_d \ll -|t_d|$, $\Delta n_1 \to 1, \Delta n_2 \to 0$. 
\end{enumerate}

Note that there is an asymmetry to the electron transfer process described above.  
In our model, only the impurity site 1 is coupled directly to the metal so that, if the energy levels of the site 1 and the site 2 are not resonant,
the site 2 is coupled to the metal only indirectly and that indirect hybridization coupling (i.e. a superexchange matrix element) will  be very small.
 Thus, within scenario (2) above, when the energy of the impurity site 2 
is far higher in energy  than  the site 1 (so that the first electron will be extracted from the site 2 and not the site 1), 
the change in the impurity population on the site 2 as a function of $\epsilon_d$ will look like a step function.  
Nevertheless, in all cases, we note that CI(N-1,N+1) remains very accurate.

\subsection{A Picture of Electron Transfer In Terms of Orbitals}

\begin{figure}[htbp]
    \centering
    \vspace*{-20mm}\hspace*{-120mm}\includegraphics[width=2.5\linewidth]{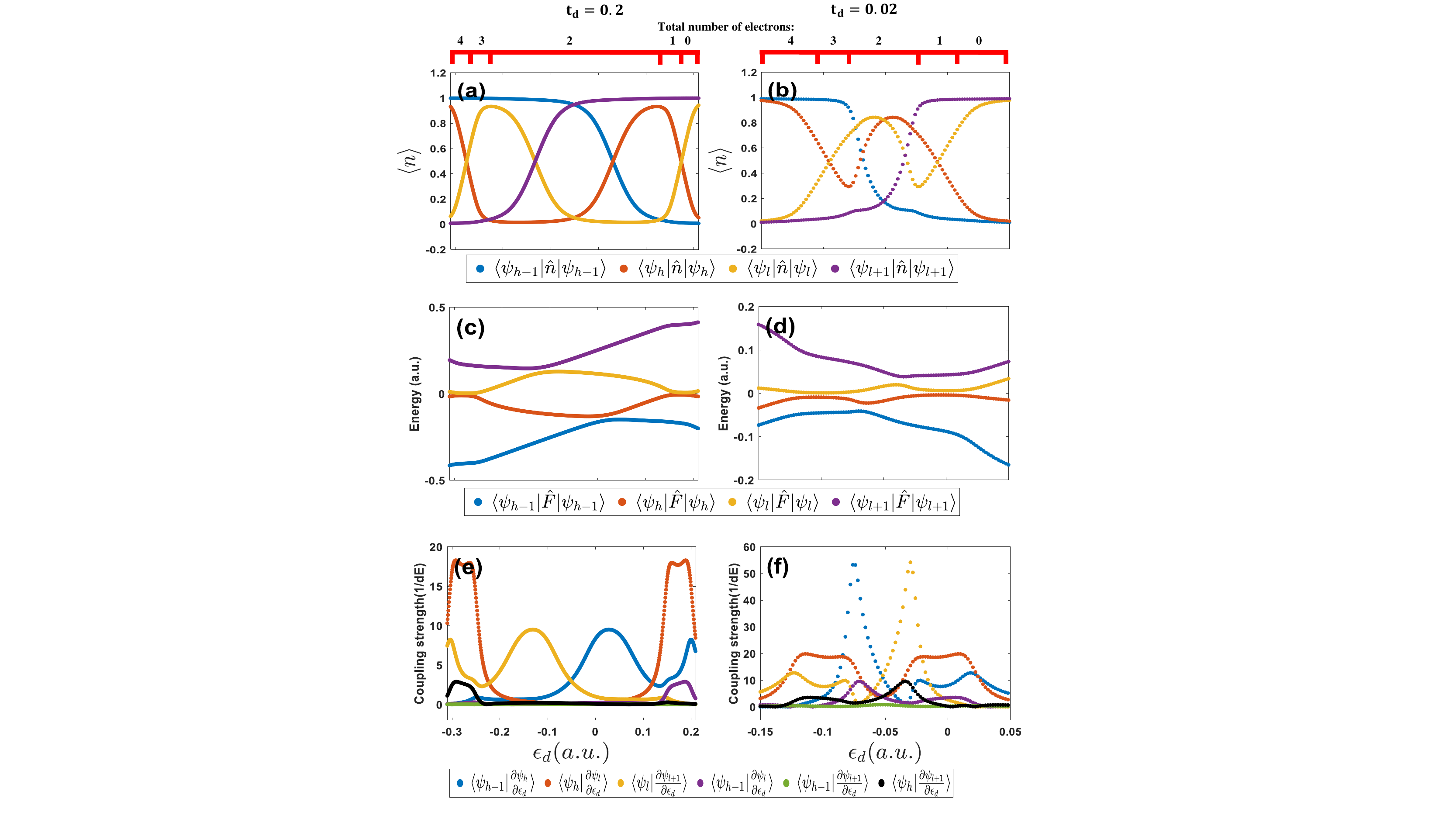}
    \vspace*{-10mm}\caption{(a-b) Impurity population of occupied entangled orbitals(OEOs) and virtual entangled orbitals(VEOs) for (a) $t_d=0.2$ and (b) $t_d=0.02$. (c-d) Orbital energies of OEOs and VEOs for (c) $t_d=0.2$ and (d) $t_d=0.02$. (e-f) Derivative couplings between OEOs and VEOs (e) $t_d=0.2$ and (f) $t_d=0.02$. Note that HOMO-1/LUMO+1 mixing remains zero for all $\epsilon_d$, so we don't include the configuration $\ket{\Phi_{h-1\overline{h-1}}^{l+1\overline{l+1}}}$ into the calculation.}
    \label{fig:orbitals}
\end{figure}

Above, we have shown that, for the case $t_d = 0.2$, CI(N-1,N-1) is applicable and can offer a reasonably accurate level of theory in terms of impurity population; however, for the case $t_d = 0.02$, CI(N-1,N+1) is necessary. At this point, it is worthwhile to explain the difference between these two cases, and why two extra configurations can be so important.
To do so, we will focus on the behavior of the relevant orbitals (two occupied entangled orbitals(OEOs) $\{\psi_{h-1},\psi_h\}$ and two virtual entangled orbitals(VEOs) $\{\psi_l,\psi_{l+1}\}$). (For a discussion of the behavior of the relevant configurations, see the S.I.)

In Fig. \ref{fig:orbitals}, we plot the impurity populations (a-b) and  energies (c-d) 
of the entangled orbitals; in (e-f), we plot the derivative couplings between
the entangled orbitals, which highlights how these orbitals change as a function of energy $\epsilon_d$ (or really
 as a function of some abstract nuclear coordinate).

We begin with the $t_d=0.2$ case and we focus on how electrons move from 
the impurity to the bath in the region $\epsilon_d\in[-0.31, -0.24]$. 
In this region,  according to Fig. \ref{fig:orbitals}c, 
there is a crossing  between the HOMO and the LUMO; and according to Fig. \ref{fig:orbitals}a,
one can ascertain that one of these orbitals
is localized on the impurity, one is delocalized in the bath, so that their crossing carries
the information about charge transfer (when the impurity moves from a charge state of -4 to -2). For this reason, one would predict that the adiabatic ground state should be compose of primarily $\{\ket{\Phi_{\textrm{HF}}},$ $\ket{S_h^{l}}, \ket{\Phi_{h\overline{h}}^{l\overline{l}}}\}$ and a CAS (2,2) calculation should be able to offer a meaningful correction to the HF solution.

Notice, however,  that the HOMO-1 and the LUMO+1 orbitals do not cross with any other orbitals in 
this energy window:  one can see that 
the HOMO-1 crosses with the HOMO  at $\epsilon_d=0.02$ (and the LUMO+1 crosses with the LUMO
at $\epsilon_d=-0.1$), and these $\epsilon_d$ values  are well within the plateau region where
the impurity has a relatively constant charge of -2. Quantitatively,
from Fig. \ref{fig:orbitals}e, we notice that the derivative couplings  between the LUMO and the LUMO+1 is centered at
$\epsilon_d=-0.1$ and is well separated from the center of the derivative couplings between the HOMO and the LUMO (which is centered at $\epsilon_d=-0.25$).  Apparently, for this value of $t_d$, the HOMO-1 and the LUMO+1 do not play a very large role in modulating the charge transfer between the HOMO and the LUMO and predicting impurity populations.
Nevertheless,  the mixing of HOMO-1/HOMO and LUMO+1/LUMO does explain why
the configurations $\{\ket{\Phi_{h\overline{h}}^{l+1\overline{l+1}}}, \ket{\Phi_{h-1\overline{h-1}}^{l\overline{l}}}\}$ are necessary to describe electron-electron correlation {\em quantitatively}, as shown in Fig. \ref{fig:n_td02}(e-f).

Next, we turn to the case $t_d = 0.02$. For this case, 
the impurity changes charge from -4 to -2 over the region
$\epsilon_d\in[-0.15, -0.08]$. Within this range, according to Figs. \ref{fig:orbitals}(b,d),
we now find two crossings:
one crossing between the HOMO and the LUMO (similar the case of $t_d=0.2$)
and another crossing  between the HOMO-1 and the HOMO 
(which is not similar the case of $t_d=0.2$). Moreover, unlike the $t_d=0.2$ case,
the HOMO-1 is not always localized to the impurity. Thus, both the HOMO and the HOMO-1 will contribute to the total two-electron transfer (one electron transferred in each step). 
This point is made even clearer when we look at the derivative couplings in Fig. \ref{fig:orbitals}f.
Here, we find that the derivative coupling between the HOMO and the HOMO-1 overlaps with the derivative coupling between the HOMO and the LUMO, highlighting the fact that one cannot fully separate
the charge transfer event ($-3 \rightarrow -2$) into two individual orbitals.
For this reason, it is not surprising that, in order to obtain an accurate description
of charge transfer, we must include the configurations: $\ket{\Phi_{h\overline{h}}^{l+1\overline{l+1}}}$ and $\ket{\Phi_{h-1\overline{h-1}}^{l\overline{l}}}$.

\newpage

\section{Conclusion and Future Directions}
\label{conclusion}

\hspace{5mm} In conclusion, we have studied the two-site Anderson impurity model problem representing a multi-electronic molecule sitting near a metal surface. After comparing the impurity population results for different CI methods, we find that CI(N-1,N-1) and CI(N-1,N+1) results match with the exact NRG results very well. Moreover, as far as the total energy is concerned, CI(N-1,N-1) (which is a relatively small CI matrix) often recovers more correlation energy than does CI(N$_\textrm{ov}$,1) (which is a relatively large CI matrix) – this statement holds rigorously in the plateau region where the impurities have 3 electrons. This finding highlights the importance of corrections from doubly excited configurations. Another key conclusion is that CI(N-1,N-1) and CI(N-1,N+1) will differ strongly in the small limit of $t_d=0.02$, where an {\em open shell singlet} can appear with 2 electrons on the impurities. In this limit, only CI(N-1,N+1) recovers correlation effects on the impurity population very well; furthermore, the method works in a robust fashion for all regimes tested so far (in terms of the intramolecular coupling strength $t_d$ and the on-site energy difference between the impurity site 2 and the impurity site 1 $\Delta \epsilon_d \equiv \epsilon_{d2} - \epsilon_{d1}$).  Clearly, when describing electron-electron interactions for a molecule on a metal surface, accurate approximations are possible (though we still need to learn more about which approximation to choose and when).  

Now, considering the minimal cost of a CAS(2,2) calculation and the moderate cost of a small CI calculation, combined with the possibility for quite reasonable accuracy when describing an impurity, the next step is to apply the present study within an {\em ab initio} DFT framework. Given that DFT can account reasonably well for dynamic correlation,
one would hope that combining DFT with configuration interaction methods should describe both dynamical correlation and static correlation. Indeed, modern density functional theory is making progress as far as calculating excited state properties using the Jacob's ladder of DFT (LSD, GGA, meta-GGA, hyper-GGA and generalized RPA) \cite{perdew2005prescription}. Thus, in the end, if DFT can be successfully merged with CI methods at a {\em metal-molecule interface} in a stable and efficient  manner (and retain accuracy), there is the exciting possibility of simulating adiabatic and non-adiabatic chemical reaction processes near metal surfaces, including charge transfer processes, bond making processes and bond breaking processes.
\begin{acknowledgement}

This work was supported by the U.S. Air Force Office of Scientific Research (USAFOSR) under Grant Nos. FA9550-18-1-0497 and FA9550-18-1-0420. We also thank the DoD High Performance Computing Modernization Program for computer time.

\end{acknowledgement}

\begin{suppinfo}

Supporting information is available online.

\end{suppinfo}

\bibliography{achemso-demo}

\begin{mcitethebibliography}{70}
\bibitem[Inp()]{Inprincipleinfinite}
In principle, of course, for an exact calculation in the condensed phase, the
  number of virtual orbitals should be infinite; nevertheless, for the present
  calculations, we let N be the total number of atomic basis functions.\relax
\mciteBstWouldAddEndPunctfalse
\mciteSetBstMidEndSepPunct{\mcitedefaultmidpunct}
{}{\mcitedefaultseppunct}\relax
\EndOfBibitem
\bibitem[Res()]{Results}
Within the results section, all figures are plotted with the parameter
  settings: the impurity on-site coulomb repulsion $U=0.1$, the hybridization
  width $\Gamma=0.01$, the energy spacing for the bath states $\Delta E=0.001$
  (or the density of the bath states $\rho(\epsilon) = 1000$), the relative
  energy difference between two impurity sites $\Delta\epsilon=0$ and the
  hopping strength between two impurity sites $t_d=0.2$.\relax
\mciteBstWouldAddEndPunctfalse
\mciteSetBstMidEndSepPunct{\mcitedefaultmidpunct}
{}{\mcitedefaultseppunct}\relax
\EndOfBibitem
\bibitem[NRG()]{NRGenergy}
Note that, we do not report exact NRG results, as the NRG procedure produces a
  total energy that depends on the discretization procedure and is not
  meaningful.\relax
\mciteBstWouldAddEndPunctfalse
\mciteSetBstMidEndSepPunct{\mcitedefaultmidpunct}
{}{\mcitedefaultseppunct}\relax
\EndOfBibitem
\end{mcitethebibliography}


\end{document}